\documentclass[aps, prd, reprint, superscriptaddress, nofootinbib, bibnotes]{revtex4-2}
\usepackage[caption=false]{subfig}
\usepackage{graphicx}
\usepackage{bm}
\usepackage{float}
\usepackage{lipsum}
\usepackage{xcolor}
\usepackage{textcomp}
\usepackage{latexsym}
\usepackage[colorlinks=true, citecolor=red,urlcolor=red]{hyperref}

\usepackage{mathtools}
\usepackage{url}
\usepackage{comment}
\usepackage{xspace}
\usepackage{acronym}
\usepackage{resizegather}
\usepackage[T1]{fontenc}
\usepackage{ae,aecompl}
\usepackage[utf8]{inputenc}
\usepackage{amssymb,amsmath,placeins,epsfig,array,bm}
\newcommand{\bol}[1]{\boldsymbol{#1}}

\newcommand{\Planck}{\textit{Planck}}

\newcommand{\Dl}{$D_{\ell}^\mathrm{TT}$}

\newcommand{\EDE}{\mathrm{EDE}}

\begin{document}

\title{$\Lambda$CDM and early dark energy in latent space:\\a data-driven parametrization of the CMB temperature power spectrum}

\author{Davide Piras}
\email[]{davide.piras@unige.ch}
\affiliation{Centre Universitaire d’Informatique, Université de Genève, 7 route de Drize, 1227 Genève, Switzerland}

\author{Laura Herold}
\email[]{lherold@jhu.edu}
\affiliation{William H. Miller III Department of Physics and Astronomy, Johns Hopkins University,
3400 North Charles Street, Baltimore, Maryland 21218, USA}

\author{Luisa Lucie-Smith}
\email[]{luisa.lucie-smith@uni-hamburg.de}
\affiliation{Max-Planck-Institut f{\"u}r Astrophysik, Karl-Schwarzschild-Str. 1, 85748 Garching, Germany}
\affiliation{Universit{\"a}t Hamburg, Hamburger Sternwarte, Gojenbergsweg 112, D-21029 Hamburg, Germany}
\author{Eiichiro Komatsu}
\affiliation{Max-Planck-Institut f{\"u}r Astrophysik, Karl-Schwarzschild-Str. 1, 85748 Garching, Germany}
\affiliation{Ludwig-Maximilians-Universit\"{a}t M\"{u}nchen, Schellingstr. 4, 80799 M\"{u}nchen, Germany}
\affiliation{Kavli Institute for the Physics and Mathematics of the Universe (Kavli IPMU, WPI), UTIAS, The University of Tokyo, Chiba, 277-8583, Japan}

\date{\today}

\begin{abstract}
Finding the best parametrization for cosmological models in the absence of first-principle theories is an open question.
We propose a data-driven parametrization of cosmological models given by the disentangled `latent' representation of a variational autoencoder (VAE) trained to compress cosmic microwave background (CMB) temperature power spectra. We consider a broad range of $\Lambda$CDM and beyond-$\Lambda$CDM cosmologies with an additional early dark energy (EDE) component. We show that these spectra can be compressed into 5 ($\Lambda$CDM) or 8 (EDE) independent latent parameters, as expected when using temperature power spectra alone, and which reconstruct spectra at an accuracy well within the \Planck{} errors. These latent parameters have a physical interpretation in terms of well-known features of the CMB temperature spectrum: these include the position, height and even-odd modulation of the acoustic peaks, as well as the gravitational lensing effect. The VAE also discovers one latent parameter which entirely isolates the EDE effects from those related to $\Lambda$CDM parameters, thus revealing a previously unknown degree of freedom in the CMB temperature power spectrum.
We further showcase how to place constraints on the latent parameters using \Planck{} data as typically done for cosmological parameters, obtaining latent values consistent with previous $\Lambda$CDM and EDE cosmological constraints. Our work demonstrates the potential of a data-driven reformulation of current beyond-$\Lambda$CDM phenomenological models into the independent degrees of freedom to which the data observables are sensitive.
\end{abstract}

\maketitle

\section{Introduction}
\label{sec:intro}

The improvement in cosmological data has allowed to determine the six parameters of the standard $\Lambda$-cold-dark-matter model ($\Lambda$CDM) to ever-increasing precision~\cite{Komatsu:2014ioa, Planck:2018vyg}, leading to the emergence of parameter tensions, for example the Hubble (or $H_0$) tension \citep[e.g.][]{Planck:2018vyg, Riess:2021jrx}. These parameter tensions have motivated the development of a plethora of alternative cosmological models, which commonly have a nested parameter structure of the form `$\Lambda$CDM + $X$', where $X$ can include additional particles or interactions. 
One model with such nested parameter structure is the early dark energy model (EDE, see Refs.~\cite{Kamionkowski:2022pkx, Poulin:2023lkg} for reviews), which is one of the most studied proposed solutions to the Hubble tension. EDE introduces three extra parameters compared to $\Lambda$CDM, whose parameter structure can give rise to so-called prior volume effects, i.e.\ upweighting of regions with larger prior volume in a Bayesian analysis, which is often unwanted~\cite{Murgia:2020ryi, Smith:2020rxx, Niedermann:2020dwg, Herold:2021ksg}. 

The parametrizations of both the $\Lambda$CDM and EDE models are motivated by human readability and theoretical considerations, giving us an intuitive understanding of the parameters of the model: the physical energy densities in CDM, $\omega_\mathrm{cdm}$, and baryons, $\omega_\mathrm{b}$, the Hubble parameter $H_0$ (or alternatively the sound-horizon size $\theta_\mathrm{s}$), the amplitude, $A_{\mathrm{s}}$, and spectral index, $n_{\mathrm{s}}$, of the primordial power spectrum, the optical depth to reionization, $\tau_\mathrm{reio}$, and the EDE parameters $(f_\EDE, \theta_\mathrm{i}, z_{\rm{c}})$, further described in Sec.~\ref{sec:EDE}. While human readability allows for easier physical interpretation, it has the disadvantage that parametrizations might be inefficient for a particular data set, namely they might lead to parameter degeneracies or prior volume effects as described above. 

Here, we thus seek for a data-driven parametrization of cosmological models, which consists of parameters that are best constrained by a given dataset. This is already commonly done in the literature, for example, in the context of galaxy weak lensing data: while the dark matter fraction, $\Omega_{\mathrm{m}}$, and the amplitude of matter clustering, $\sigma_8$, show a `banana-shaped' degeneracy, the product $S_8 \equiv \sigma_8\sqrt{\Omega_{\rm{m}}/0.3}$ is well-constrained by weak lensing surveys~\cite{Heymans:2020gsg, Kilo-DegreeSurvey:2023gfr, Sugiyama:2023fzm}.

In this work, we search for a data-driven parametrization of cosmic microwave background (CMB) temperature (TT) power spectra.
To do so, we make use of a variational autoencoder (VAE), a neural network architecture consisting of an encoder-decoder structure \cite{KingmaWelling2013, Rezende2014}: the encoder compresses the data (here, CMB TT power spectra) into a chosen number of \textit{latent} variables (or simply latents), which in turn are used by the decoder to reconstruct the data. The latents obtained by a trained VAE represent a data-optimized parametrization of the CMB spectra. By training the VAE with two sets of CMB TT spectra, one set generated assuming $\Lambda$CDM and one assuming EDE as the underlying cosmological model, we can obtain alternative data-driven parametrizations of these two models. 

A related approach was pursued in Ref.~\cite{Hu_2001}, where it was shown that measurements of the CMB power spectrum can be understood in terms of a phenomenological representation of four variables. Our approach is similar, except that our reparametrization of the CMB features (and therefore of the underlying cosmological parameters) is performed by a neural network and is thus entirely data-driven, as well as non-linear. Refs.~\cite{Kosowsky:2002zt, Jimenez:2004ct} additionally explored physical parameter combinations that break degeneracies and accelerate the computation of CMB spectra, Refs.~\cite{Huterer:2002hy, Crittenden:2005wj, Zhao:2009fn, Hojjati:2011xd, Asaba:2013xql} investigated data-driven parametrizations of cosmological models based on principal component analysis (PCA), a linear compression scheme, while Ref.~\cite{Piras:2023lid} used of a VAE to compress $w$CDM cosmologies using matter power spectra. 

Our goal in this work is to answer three questions:
\begin{enumerate}
    \item Into how many (latent) parameters can the CMB power spectra be compressed while still retaining high predictive accuracy, i.e.\ will we recover the same number of parameters as in $\Lambda$CDM or EDE, respectively, or fewer? 
    \item Do the data-driven parametrizations represent known cosmological parameters or effects, i.e.\ does the neural network recover human-interpretable parameters?
    \item Can we obtain meaningful constraints on the latent parameters using real CMB data?
\end{enumerate} 
Answering these questions paves the way towards the use of data-driven parametrizations for inference in cosmology, and possibly address the limitations given by prior volume effects in cosmological inference. 

The paper is structured as follows. In Sec.~\ref{sec:EDE} we briefly review the EDE model, while in Sec.~\ref{sec:method} we describe the data and our methodology. In Sec.~\ref{sec:results} we present the results in terms of accuracy of the reconstructed spectra and constraints on the latent parameters, while Sec.~\ref{sec:interpret} focuses on the physical interpretation of the latents. We conclude in Sec.~\ref{sec:conclusions}.

\begin{figure*}
\centering
\includegraphics[trim={0.cm 12cm 0.35cm 0
cm},width=\textwidth]{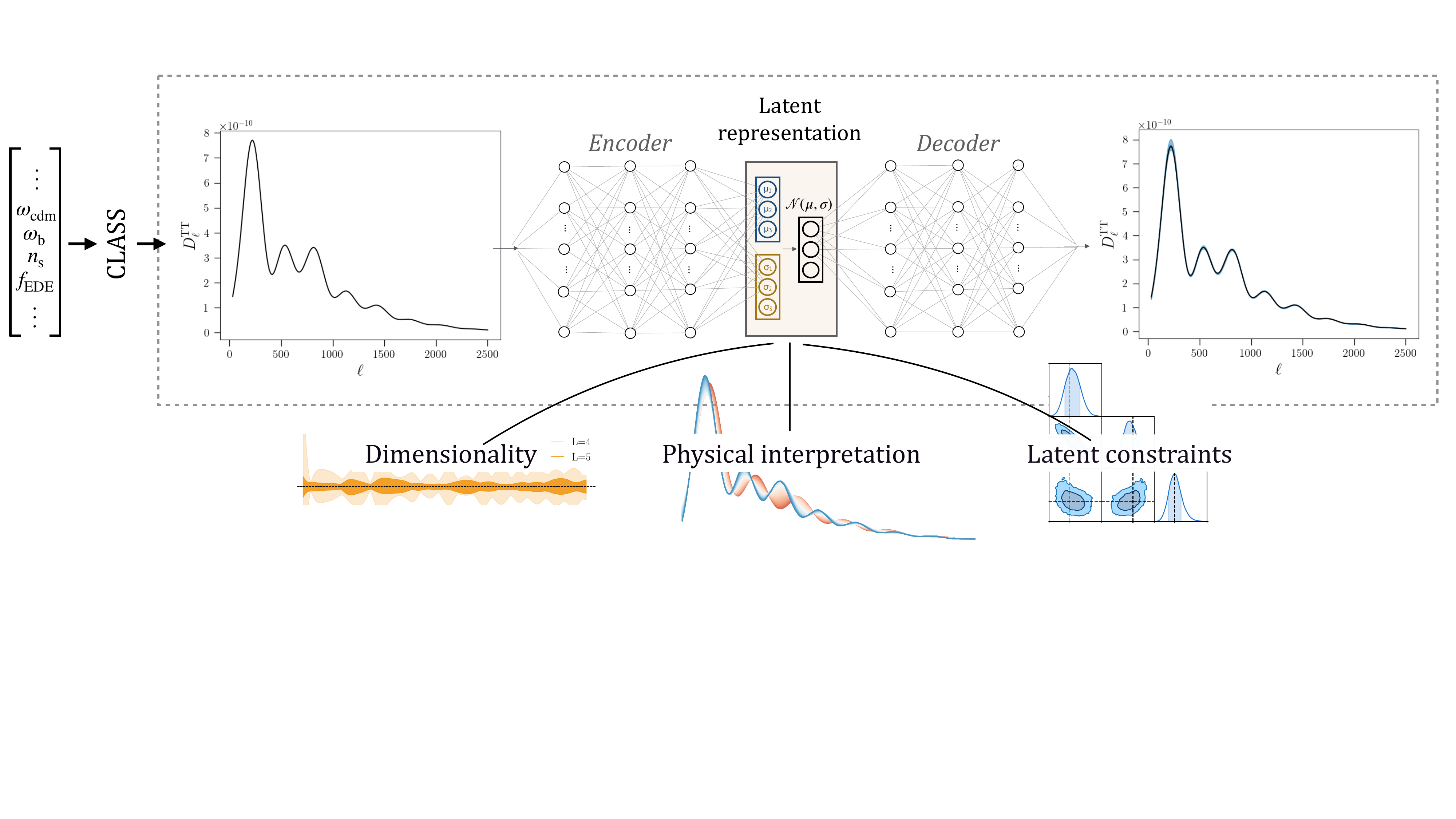}
\caption{Our method consists of a variational autoencoder (VAE), which compresses the CMB temperature power spectrum into a low-dimensional latent representation (via the \textit{encoder}); the representation is then sampled to reconstruct CMB spectra (via the \textit{decoder}). Our goal is to (i) find the minimum number of latents required to reconstruct accurate spectra, (ii) physically interpret the information captured by the latents, and (iii) provide constraints in latent space using \Planck{} data and relate them to latents for different cosmological models. 
}
\label{fig:illustration}
\end{figure*}

\section{Early dark energy}
\label{sec:EDE}
EDE denotes a class of models which feature a dark energy-like growth in the early Universe, but become subdominant after recombination, $z^*$. The boost of expansion rate in the early Universe, $H(z)$, leads to a reduction of the physical size of the sound horizon, $r_{\rm{s}}(z^*) = \int_{z^*}^\infty c_{\rm{s}}(z)/H(z)\mathrm{d}z$, where $c_{\rm{s}}(z)$ is the sound speed in the baryon-photon plasma. Fixing the angular size of the sound horizon, $\theta_{\rm{s}} = r_{\rm{s}}/D_{\rm{A}}$, which is directly and precisely measured by CMB observations, translates into a reduction of the angular diameter distance to recombination, $D_{\rm{A}}(z^*) = \int_0^{z^*} \mathrm{d}z/H(z)$, which in turn requires an increase in the Hubble parameter, $H_0$, alleviating the Hubble tension~\cite{Kamionkowski:2022pkx, Poulin:2023lkg}. 

The most commonly studied EDE model is the axion-like EDE model~\cite{Karwal:2016vyq, Poulin:2018dzj, Poulin:2018cxd}, which introduces a scalar field $\phi$ with potential $V(\phi) = V_0 [1-\cos(\phi/f)]^n$, where $V_0 =~m^2f^2$ with $m$ and $f$ being referred to as `mass' and `decay constant', respectively. The index $n$ is typically fixed to $n=3$ as this presents the best fit to data \cite{Smith:2019ihp, Poulin:2023lkg}. For parameter inference, these three parameters are commonly translated into the `phenomenological parameters': $f_\EDE = \rho_{\rm EDE}(z_{\rm{c}})/\rho_{\rm tot}(z_{\rm{c}})$, the maximum fraction of EDE; $z_{\rm{c}}$, the `critical redshift' at which the EDE field starts to oscillate in its potential; and $\theta_\mathrm{i}$, the initial value of the scalar field in the potential.

Analyses of the EDE model including \Planck\ CMB data~\cite{Planck:2018vyg} and large-scale structure (LSS) data along with the direct measurement of $H_0$ by the SH0ES collaboration~\cite{Riess:2021jrx} indicate a promising ability of EDE to resolve the Hubble tension \cite{Poulin:2018cxd, Smith:2019ihp, Schoneberg:2021qvd}. However, excluding direct measurements of $H_0$ from the analysis generally leads to tight upper limits on the fractions of EDE, $f_\EDE$, and lower values of $H_0$~\cite{Murgia:2020ryi, Hill:2020osr, Ivanov:2020ril, DAmico:2020ods, Smith:2020rxx, McDonough:2021pdg, Gsponer:2023wpm}, challenging the ability of EDE to resolve the tension. These tight upper limits on $f_\EDE$ are partially driven by prior volume (or projection) effects in the Markov chain Monte Carlo (MCMC) posterior, which arise due to the complicated nested parameter structure of the model: while $f_\mathrm{EDE}$ controls the fraction of EDE, $z_{\rm{c}}$ and $\theta_\mathrm{i}$ are auxiliary parameters encoding details of the model. When $f_\mathrm{EDE}$ approaches zero, the $\Lambda$CDM limit is recovered: both $z_{\rm{c}}$ and $\theta_\mathrm{i}$ become then redundant and unconstrained. This leads to a larger prior volume in $f_\mathrm{EDE} \approx 0$ than $f_\mathrm{EDE} > 0$ and a non-Gaussian posterior, which in turn can lead to a preference for the $\Lambda$CDM limit in the marginalized posterior \cite{Murgia:2020ryi, Smith:2020rxx, Niedermann:2020dwg}. This is backed up by frequentist analyses of the EDE model using profile likelihoods, which show a preference for large fractions of EDE and values of $H_0$ in agreement with the direct measurements \cite{Herold:2021ksg, Herold:2022iib, Reeves:2022aoi, Gomez-Valent:2022hkb}.\footnote{Replacing the \Planck\ baseline CMB data with alternative CMB data \cite{Hill:2021yec,Poulin:2021bjr,Efstathiou:2023fbn, McDonough:2023qcu} changes this simplified story.}

While there are more and more challenges for the axion-like EDE model, e.g.\ a worsening of the $S_8$ tension in EDE cosmologies~\cite{Smith:2019ihp, Hill:2020osr, Ivanov:2020ril, Vagnozzi:2021gjh, Ye:2021nej, Pedreira:2023qqt} and inability to fit certain CMB and LSS data sets~\cite{Efstathiou:2023fbn, McDonough:2023qcu, Goldstein:2023gnw}, EDE remains a promising class of models and interesting test case for solutions to the Hubble tension.

\section{Overview of the method}
\label{sec:method}

An illustration of our method is shown in Fig.~\ref{fig:illustration}. We first generate theoretical predictions for the unbinned data vector consisting of CMB temperature power spectra $D_{\ell}^{\mathrm{TT}}$, using the Einstein-Boltzmann solvers \texttt{CLASS} \citep{Lesgourgues2011, Blas2011} or \texttt{CLASS\_EDE} \cite{Hill:2020osr}, considering a broad range of cosmological parameters for both a $\Lambda$CDM or EDE model. We then train a $\beta$-VAE \cite{Higgins17}, a regularized version of a VAE \cite{KingmaWelling2013, Rezende2014}, to (i) compress the information contained in a $D_{\ell}^{\mathrm{TT}}$ spectrum into an $L$-dimensional Gaussian latent representation (via the \textit{encoder}), and (ii) reconstruct the spectrum from samples in latent space via the \textit{decoder}. In Fig.~\ref{fig:illustration}, we considered a 3D latent space for visualization purposes. 

We use the trained VAE model to discover the underlying dimensionality of the CMB temperature power spectrum, interpret the information contained in the latent space, and provide posterior constraints on the latent parameters. The minimal number of latent variables required to describe the data is found through an iterative process: we increase the latent dimensionality iteratively until we find the lowest number of latents such that the reconstruction accuracy is well within the 1$\sigma$ error from \Planck{}. The latents' physical interpretation is achieved through the inspection of latent traversals and the use of mutual information, a well-known information-theoretic metric which we describe in more detail in Sec.~\ref{sec:MI_calculation}. We also obtain constraints of the latent parameters using the trained decoder and the \Planck{} data via an MCMC analysis, showcasing that it is possible to obtain meaningful posterior contours for this data-driven parametrization. We describe each of these steps in more detail in the next sections. In this paper, `$\log$' denotes the decimal logarithm, while `$\ln$' the natural logarithm.

\subsection{Training data: theoretical predictions for \Dl} \label{sec:DlTT_class}
We construct a Latin hypercube of cosmological parameters in order to generate theoretical predictions for the CMB temperature power spectrum. 
This is performed for two cosmological scenarios: the standard $\Lambda$CDM model and an extended model incorporating EDE, denoted simply as EDE.
The $\Lambda$CDM model includes six standard parameters ($\omega_{\mathrm{b}}$, $\omega_{\rm cdm}$, $h=H_0/100$, $\tau_{\rm reio}$, $n_{\rm s}$, $\ln 10^{10} A_{\rm s}$), while for the EDE cosmology there are three additional cosmological parameters ($f_{\rm EDE}$, $\theta_\mathrm{i}$, $\log z_{\rm c}$), as defined in Sec.~\ref{sec:EDE}.

The parameter ranges are reported in Table~\ref{tab:prior_lh}, and span several standard deviations around the \Planck{} best-fit parameters~\cite{Planck:2018vyg}, thus also covering the SH0ES results on $h$~\cite{Riess:2021jrx}. We choose a lower bound $f_{\rm EDE} = 0$ unlike other literature on EDE (e.g.\ \citep{Hill:2020osr, Reeves:2022aoi}) to include the $\Lambda$CDM case in the EDE analysis: while there are no instances in the training set where $f_{\rm EDE}$ is exactly equal to 0, as this value represents the lower bound of the Latin hypercube sampling, we have confirmed that the CMB temperature power spectra for $\Lambda$CDM and $\Lambda$CDM+$f_{\rm EDE}$=0.001 cosmologies exhibit a fractional difference of less than $0.05\%$ on average. This indicates that small values of $f_{\rm EDE}$ produce spectra that are effectively indistinguishable from those with $f_{\rm EDE} = 0$, implying that the VAE trained on EDE spectra encounters examples of $\Lambda$CDM-like spectra during training. Being able to extend the prior all the way down to $f_{\rm EDE}=0$ is an improvement over standard MCMC analyses on cosmological parameters, which often impose a lower prior $f_{\rm EDE}>0$ in order to minimize prior volume effects.

Given a set of cosmological parameters, we use the publicly available Einstein-Boltzmann solvers \texttt{CLASS} and \texttt{CLASS\_EDE}, where the latter is an extension of the former which includes EDE.
We use these Einstein-Boltzmann solvers to generate the theoretical CMB temperature power spectrum $D_{\ell}^{\mathrm{TT}}$ in the range $\ell \in [30, 2500]$, which is covered by the \texttt{Plik\_lite} likelihood \cite{Planck:2019nip, Prince:2019hse};\footnote{\href{https://github.com/heatherprince/planck-lite-py}{https://github.com/heatherprince/planck-lite-py} the highest multipole considered by \texttt{Plik\_lite} is 2508, therefore we discard the last $\ell$ bin.} we use this likelihood throughout our analysis. All other \texttt{CLASS}-related parameters are left to their standard values.
The training data, comprised of CMB spectra from the standard $\Lambda$CDM model and the EDE model, are utilized to train two VAEs independently, which we denote VAE$_{\Lambda \mathrm{CDM}}$ and VAE$_{\mathrm{EDE}}$, respectively.

We create 500\,000 $D_{\ell}^{\rm TT}$ spectra, and use $80\%$ of spectra for training, $10\%$ for validation and the rest for testing. Before feeding these spectra as input to the VAE, in order to facilitate training we divide the spectra by a reference spectrum (different for the two VAEs, but purely arbitrary), take the decimal logarithm and standardize the data. The predictions made by the VAE are then always parsed through these operations in reverse order to obtain the final $D_{\ell}^{\rm TT}$ predictions.

\begin{table}
  \centering 
  \begin{tabular}{c|c}
    \textbf{Parameter}               &  \textbf{Prior range}   \\
    \hline
    \hline
    $\omega_{\mathrm{b}}$            &   [0.020, 0.024]    \\
    \hline
    $\omega_{\mathrm{cdm}}$          &    [0.10, 0.13]        \\
    \hline
    $h$                              &    [0.62, 0.80]         \\
    \hline
    $\tau_{\mathrm{reio}}$           &    [0.01, 0.13]         \\
    \hline
    $n_{\rm{s}}$                            &     [0.92, 1.01]      \\
    \hline
    $\mathrm{ln}10^{10}A_{\rm{s}}$          &      [2.90, 3.18]       \\
    \hline
    \hline
    $f_{\rm{EDE}}$          &      [0, 0.5]       \\
    \hline
    $\theta_\mathrm{i}$          &      [0.1, 3.1]       \\
    \hline
    $\mathrm{log} z_{\rm{c}}$          &      [3, 4.3]       \\
    \hline
    \hline
    \end{tabular}
   \caption{Prior ranges to generate the training data Latin hypercube. These priors cover 10 standard deviations around the combined \Planck{} 2018 best-fit results (rightmost column in Table 1 in Ref.~\citep{Planck:2018vyg}), except for the lower bound on $\tau_{\mathrm{reio}}$ which is taken from \texttt{CosmoPower} \citep{SpurioMancini22} (otherwise it would be negative), and the upper bound on $h$ (which would not include the SH0ES \citep{Riess:2021jrx} result otherwise). For the EDE parameters, we use commonly chosen prior ranges, e.g.\ \cite{Poulin:2023lkg}.}
  \label{tab:prior_lh}
\end{table}

\subsection{Variational autoencoders} 
VAEs are unsupervised encoder-decoder networks that learn to compress the input data into a lower-dimensional representation, known as the latent representation or latent variables, and then use these latents to reconstruct something that is closely similar to the input data \cite{Hinton1993, KingmaWelling2013, Rezende2014}. The former part of the algorithm is the \textit{encoder} and the latter the \textit{decoder}; we choose the network architectures of the encoder and decoder to be simple 1D convolutional neural networks. The latent representation aims to capture all the relevant information required to reconstruct the input. The latent representation of a given input $\bm{x}$ is a probability distribution function $p(\bm{z} | \bm{x})$ which is usually represented by a multivariate diagonal Gaussian $p(\bm{z} | \bm{x}) = \mathcal{N} (\bm{\mu}, \bm{\sigma})$, where $\bm{\mu}$ and $\bm{\sigma}$ are the means and standard deviations of the Gaussian distribution of each latent parameter $z$. The size of the vectors $\bm{\mu}$ and $\bm{\sigma}$ is $L$, namely the latent space dimensionality. The means and standard deviations of each latent dimension are the outputs of the encoder, while the decoder takes as input samples $\bm{z} \sim \mathcal{N} (\bm{\mu}, \bm{\sigma})$, thus returning a distribution of reconstructed outputs $\bm{\hat{x}}$ from a single input $\bm{x}$. A VAE can be thought of as a generalization of PCA to non-linear compressions: while PCA performs a linear decomposition of the original input space, the VAE performs a \textit{non-linear} compression via the non-linear convolutional layers of the encoder.

Typically, training a VAE involves minimizing a reconstruction loss, measuring how well the decoder can reconstruct an output that is identical to the original input, starting from the latent representation. When the latent representation allows a good reconstruction of its input, then it has retained the most important information present in the input data. In addition to the reconstruction term, we also include a regularization term in the loss function that promotes \textit{disentanglement} in latent space: that is, the independent factors of variation in the CMB temperature power spectrum are captured by different, independent latents. The loss function is then given by:
\begin{equation}
\mathcal{L} = \mathcal{L}_\mathrm{recon}(D^{\mathrm{TT}}_{\ell, \mathrm{true}}, D^{\mathrm{TT}}_{\ell, \mathrm{pred}}) + \beta \, \mathcal{D}_\mathrm{KL}[p(\bol{z} | \bol{x}); q(\bol{z})] \ ,
\label{eq:loss}
\end{equation}
 where the second term is the Kullback-Leibler (KL) divergence \citep{Kullback1951} between the latent distribution returned by the encoder $p(\bol{z} | \bol{x})$ and a prior distribution over the latent variables $q(\bol{z})$, which we take to be $\mathcal{N}(\bol{0}, \bol{1})$. For the first term, $\mathcal{L}_\mathrm{recon}(D^{\mathrm{TT}}_{\ell, \mathrm{true}}, D^{\mathrm{TT}}_{\ell, \mathrm{pred}})$, we choose the mean squared error. The parameter $\beta$ weighs the KL divergence term with respect to the predictive term, and must be carefully optimized to achieve disentanglement without significantly affecting the reconstruction accuracy. 
 A VAE with the loss function as in Eq.~\eqref{eq:loss} is usually referred to as a $\beta$-VAE \cite{Higgins17}, and the latent representation can be thought of as the independent degrees of freedom in the input.

We train the VAEs using the \texttt{Adam} optimizer \cite{Kingma15}, decreasing the learning rate by a factor of 10 between 10$^{-3}$ and 10$^{-5}$ each time the validation loss does not improve for 50 consecutive epochs, and with a batch size of 1024. After each convolutional layer, we apply batch normalization~\cite{Ioffe15} and a trainable activation function as described in Ref.~\cite{Alsing20} to increase training efficiency. Training a single model until convergence typically requires less than 24 hours, using a single GPU with up to 24 GB of memory.

\section{Results}
\label{sec:results}
\subsection{Accuracy of the reconstructed \Dl from the VAE}
\begin{figure}
\centering
\includegraphics[trim={0.9cm 0.7cm 1.1cm 0.7cm},clip, width=\linewidth]{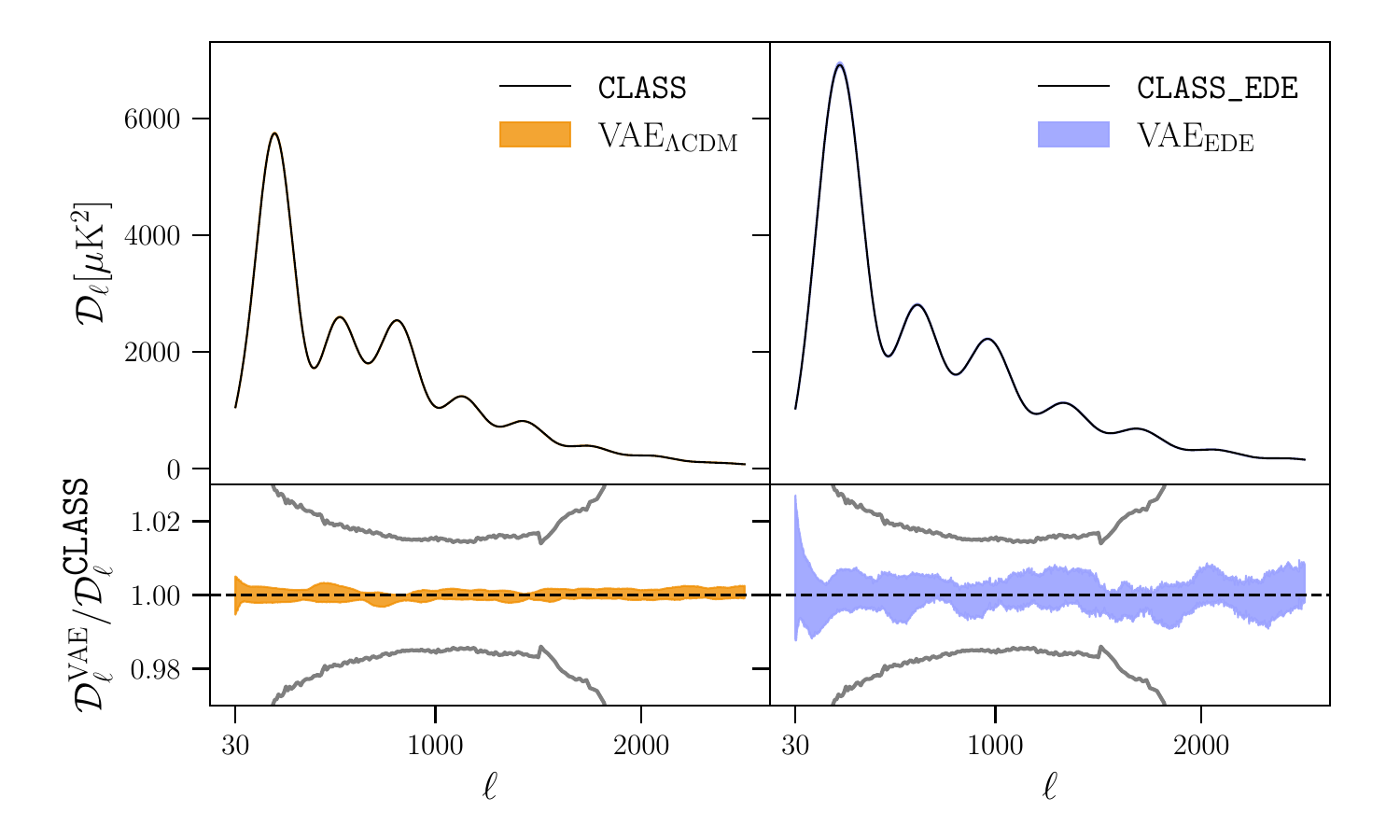}
\caption{\textit{Top panels:} Examples of \Dl{} for two different cosmologies -- a $\Lambda$CDM cosmology with the \Planck{} best-fit cosmological parameters (left panel) and an EDE one selected randomly from the test set (right panel). In black we show the spectra generated with the Einstein-Boltzmann solvers, while the shaded regions cover the range of reconstructed spectra returned by the VAE decoder given 100 samples in latent space. \textit{Bottom panels:} Ratio between the reconstructed and \texttt{CLASS} or \texttt{CLASS\_EDE} spectra. In gray, we show the \Planck{} $1\sigma$ errors for comparison.}
\label{fig:Dl_bestfit_example}
\end{figure}

Fig.~\ref{fig:Dl_bestfit_example} shows examples of the reconstructed and \texttt{CLASS} CMB temperature power spectra for two different cosmologies. In the left panel, we show the spectrum returned by \texttt{CLASS} (black line) given the best-fit $\Lambda$CDM cosmological parameters from \Planck{} \citep{Planck:2019nip}. In shaded orange, we show the reconstructed spectrum from the VAE$_{\Lambda\rm{CDM}}$ model for the same cosmology, sampling 100 times from the latent space. In the right panel, we show the spectrum returned by \texttt{CLASS\_EDE} for a random EDE test set cosmology. In shaded blue, we show the reconstructed spectrum from the VAE$_{\rm{EDE}}$ model for the same cosmology, sampling 100 times from the latent space. The VAEs return unbiased predictions with sub-percent uncertainty throughout the entire $\ell$ range; this is well within the $1\sigma$ error from \Planck{} (marginalized over nuisance parameters as in the \texttt{Plik\_lite} likelihood), shown as a gray line.\footnote{We also considered using the Simons Observatory \cite{SimonsObservatory:2018koc} forecast errors as a benchmark, but the resulting constraints are looser than or similar to \Planck{} for $\ell\lesssim 1500$.}

\begin{figure}
\centering
\includegraphics[trim={0.9cm 1.5cm 1.25cm 1cm},clip, width=\linewidth]{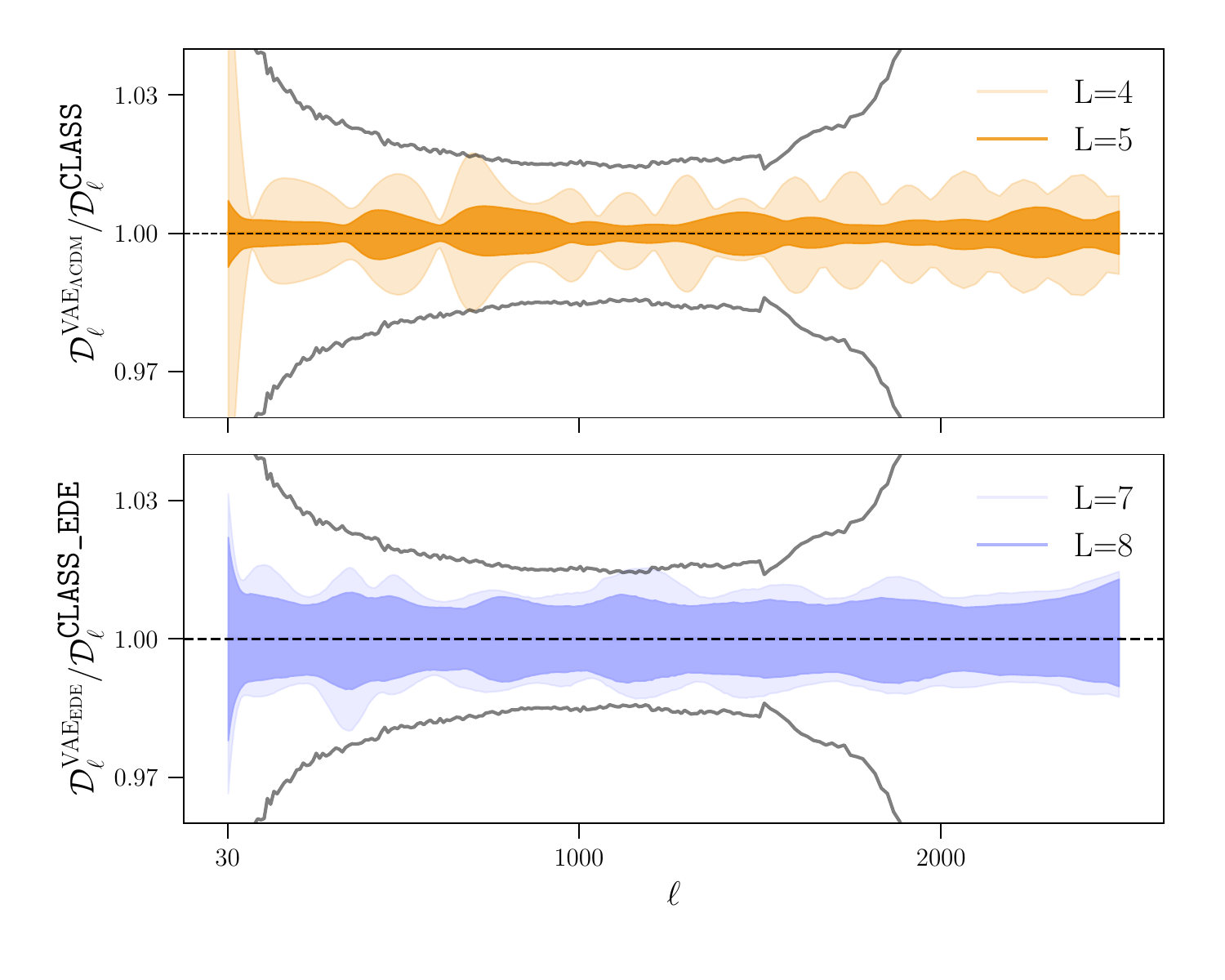}
\caption{Mean and 99\% confidence interval of the residual error for the entire test set cosmological parameter space. \textit{Top panel:} Residual error of two VAEs, both trained on $\Lambda$CDM TT spectra, with 4D and 5D latent dimensionality respectively. \textit{Bottom panel:} Same for two VAEs trained on EDE TT spectra with 7D and 8D latent dimensionality respectively. The values of $\beta$ for all these models are tuned such that we obtain disentangled latents. The \Planck{} $1\sigma$ errors are shown in gray.}
\label{fig:vae_accuracy}
\end{figure}

A more quantitative, global measure of the overall performance of the two VAEs in reconstructing $D_{\ell}^{\rm TT}$ is shown in Fig.~\ref{fig:vae_accuracy}. We take the ratio between the VAE-reconstructed and the \texttt{CLASS} spectra for each set of cosmological parameters set aside for testing the VAE, and show the mean (line) and 99\% confidence interval (shaded region) of such ratio. 
Again, the gray line indicates the 1$\sigma$ \Planck{} error, shown as a reference since we want the VAE to return predictions well within it.

In each panel, we show the performance of two disentangled VAEs trained using different latent dimensionalities $L$, denoted in the legend. In all cases, the mean residual is always consistent with 1, meaning that the VAE always returns unbiased predictions irrespective of latent dimensionality and cosmology. The variance in the residuals instead varies depending on the latent dimensionality of the specific VAE model and the value of $\beta$ in the loss function. If the latent dimensionality is too small to encode all the information present in the power spectrum, the variance will be large. Moreover, the value of $\beta$ must be set to give, for a given $L$, the best possible disentanglement without significantly increasing the variance of the model errors. 

In Fig.~\ref{fig:vae_accuracy}, we show the residuals of the models with the lowest value of $\beta$ that achieve disentanglement. For $\Lambda$CDM, we find that the best performance (meaning highest accuracy and disentanglement) is achieved with 5 latent parameters; in other words, we find that the CMB temperature power spectrum can be described by five degrees of freedom for $\Lambda$CDM cosmologies. This number is expected since the spectra are generated with six $\Lambda$CDM parameters, of which $A_{\rm{s}}$ and $\tau_\mathrm{reio}$ are degenerate as we are not including any polarization data. We thus recover the same number of degrees of freedom as in the $\Lambda$CDM parametrization. We also show the residuals of the 4-latent model for comparison; the error increases to a degree comparable to the $1\sigma$ error from \Planck{}, which therefore makes us discard this model. 

The bottom panel of Fig.~\ref{fig:vae_accuracy} shows the case for EDE cosmologies. Here, we find that an 8-dimensional latent space can achieve good enough accuracy to be well below the \Planck{} error. 
Considering $L=7$ latents increases the error slightly, to a level which becomes comparable to the \Planck{} observational error. However, we note that the difference in accuracy between the $L=7$ and $L=8$ model is small, meaning that the additional degree of freedom contributes to only a small amount of information about the \Dl{}; yet, this information is needed to fulfill our accuracy requirement. Thus, also for the EDE model we recover the expected number of degrees of freedom, i.e.\ nine $\Lambda$CDM+EDE parameters, minus one due to the $A_{\rm{s}}$-$\tau_\mathrm{reio}$ degeneracy. We also note that the VAE$_{\rm EDE}$'s accuracy is slightly worse than that of the VAE$_{\Lambda\rm{CDM}}$, which is expected due to the non-trivial contributions of EDE to the CMB TT power spectrum.

\subsection{MCMC analysis}
\begin{figure}
\centering
\includegraphics[trim={0.6cm 0.2cm 0.2cm 0.2cm},clip, width=\linewidth]{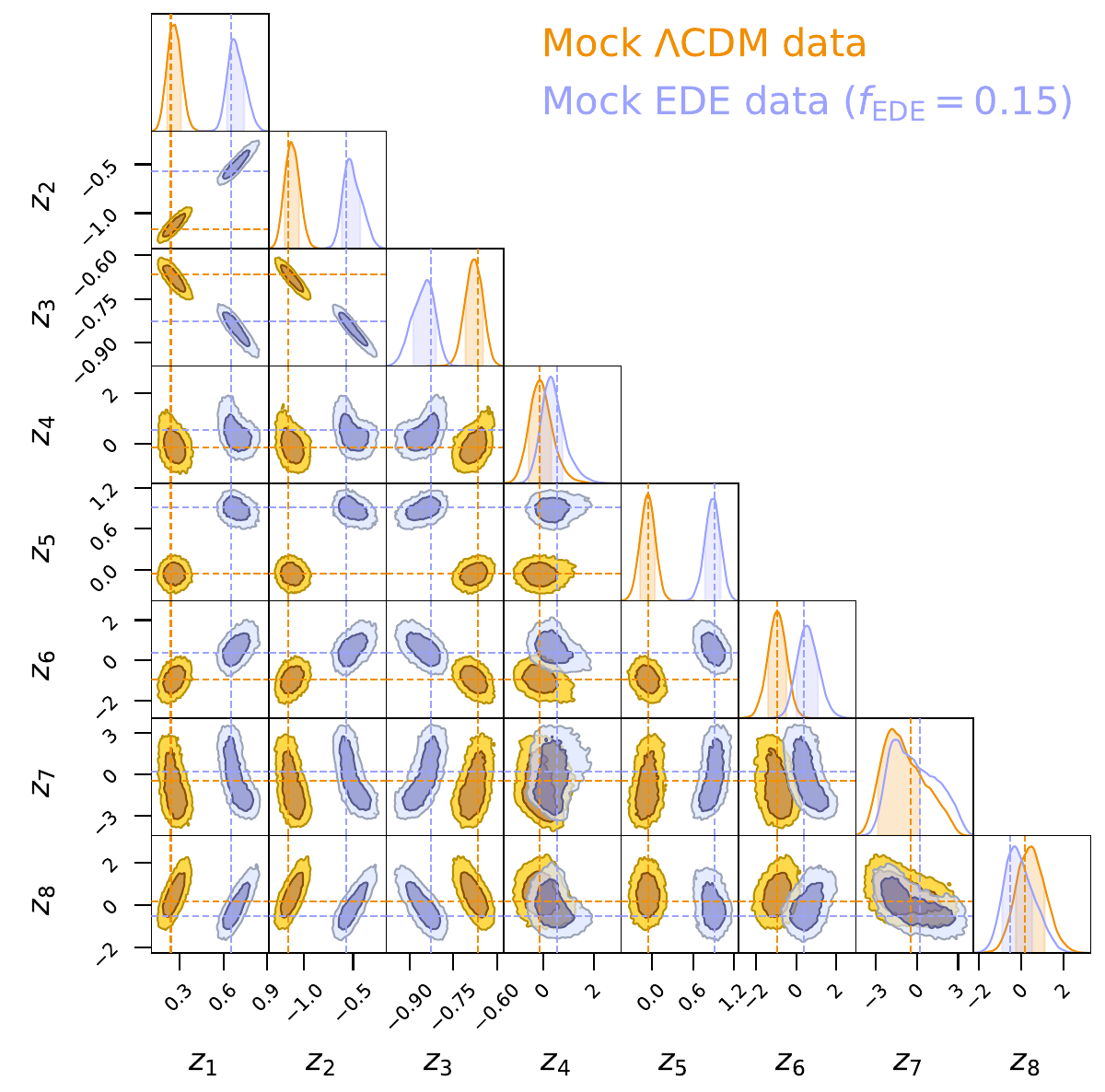}
\caption{1D and 2D marginalized posterior probability distributions for the eight latent parameters $\bm{z}$ given the two examples of mock data. The mock data are generated by the decoder given a `ground truth' point in the 8D latent space; these are marked by dashed lines. These points correspond to respectively the most likely latent values of a \Planck{} best-fit $\Lambda$CDM cosmology (orange) and an EDE model with $f_{\rm EDE}=0.15$ (blue). We show unbiased and accurate constraints in latent space, thus validating the robustness and trustworthiness of our pipeline.
}
\label{fig:mocktest}
\end{figure}

\begin{figure*}
    \centering
    \subfloat[VAE$_{\Lambda\rm{CDM}}$]{%
  \includegraphics[width=\columnwidth]{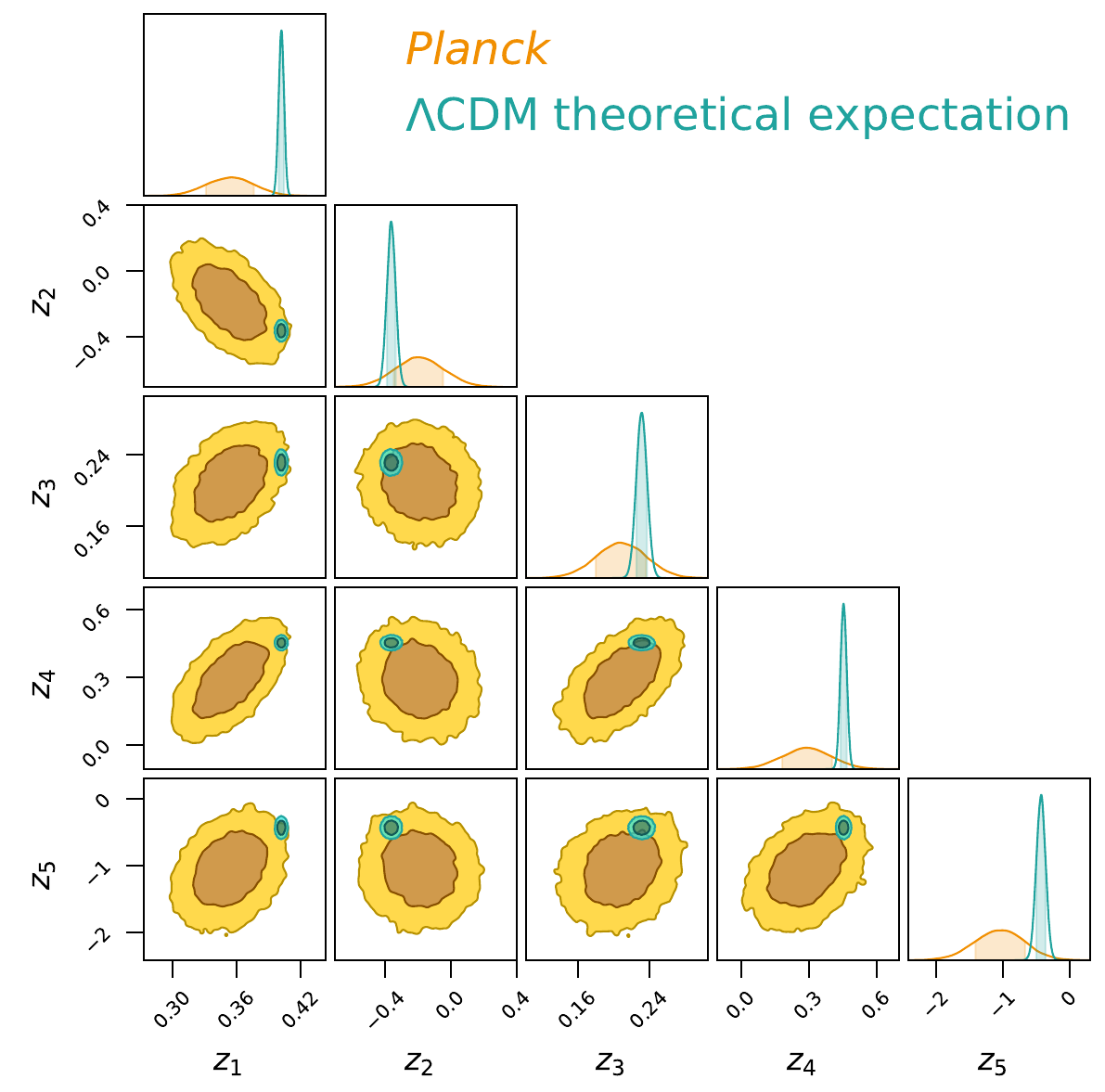}%
}
    \subfloat[VAE$_{\rm{EDE}}$]{%
  \includegraphics[width=\columnwidth]{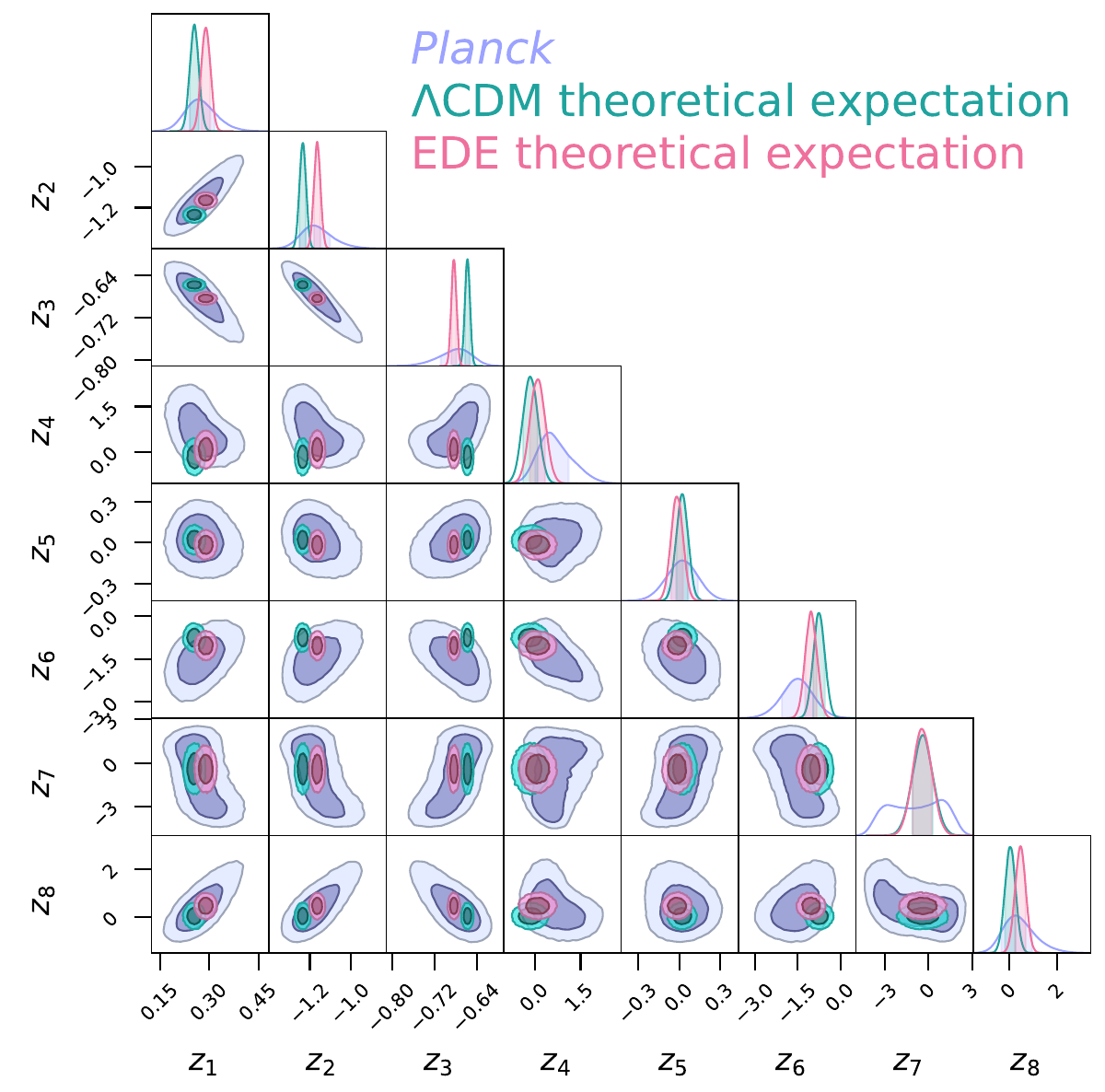}%
}
    \caption{1D and 2D marginalized posterior probability distributions for the latent parameters $\bm{z}$ given the \Planck{} data; the left panel shows the $\Lambda$CDM latent parameters in orange, and the right panel the EDE ones in blue. We compare the latent posterior constraints to theoretical expectations for the range of latent values allowed by a given set of cosmologies: a $\Lambda$CDM cosmology with best-fit parameters from Ref.~\cite{Planck:2018vyg} (green in both panels) and an EDE cosmology with best-fit parameters from \texttt{Plik\_lite} (pink in the right panel). Our constraints in latent space are thus consistent with previous constraints from the literature.
    }
    \label{fig:lcdmrefmodel_planck}
\end{figure*}

We now move onto performing parameter inference of the latents of the VAEs. We use the \texttt{emcee} sampler \cite{ForemanMackey13} to produce posterior constraints of the latent parameters using the VAE decoder model and the \Planck{} $D_{\ell}^{\mathrm{TT}}$ data vector. We adopt a uniform prior between $-$5 and 5 for each latent parameter, although we also tested sampling from a Gaussian mixture model fitted to the set of latent Gaussian distributions corresponding to the test set cosmologies, finding no significant difference in the final posterior constraints.
The \texttt{emcee} sampling is typically initialized with 64 walkers with an initial point sampled from a unit Gaussian with zero mean, and then proceeds until convergence. It typically takes about 3 hours on 12 CPUs to reach convergence, which we assess by ensuring that the number of iterations is at least 100 times the estimated autocorrelation time. We also verified that replacing \texttt{emcee} with a nested sampler does not change the results. 

For all MCMC analyses throughout this work, we use the \texttt{Plik\_lite} likelihood code to compare the theory $D_{\ell}^{\mathrm{TT}}$ generated by the VAE to the (mock or real) data. We bin the theoretical predictions returned by the decoder using the same binning scheme as for the data; the (binned) theoretical predictions can then be used as input to the \texttt{Pklik\_lite} code to estimate the likelihood, as described in Ref.~\cite{Planck:2019nip}. 

\subsubsection{Latent parameter constraints from mock data}

Before applying our pipeline to the real \Planck{} data, we perform a validation test of our approach using two mock data spectra and the trained $\rm{VAE}_{\rm{EDE}}$. The mock data were generated by the decoder given two different `ground truth' points in the 8D latent space. These points correspond respectively to the most likely latent values of a $\Lambda$CDM cosmology with best-fit values from \Planck{}, and an EDE model with $f_{\rm EDE}=0.15$, $\theta_\mathrm{i} = 2.8$, and $\log z_\mathrm{c} = 3.6$. The choices were made in order to pick two points in latent space which are distant from each other due to the presence of a significant EDE component; this choice additionally allows us to visualize how sensitive the latent space is to the $f_{\rm EDE}$ parameter. 

We run our pipeline independently for each mock $D_{\ell}^{\mathrm{TT}}$ data and show the 1D and 2D marginalized posterior probability distributions of the latent parameters in Fig.~\ref{fig:mocktest}. The `ground truth' latent parameters used to generate the mock data are marked by dashed lines, one for the \Planck{} best-fit $\Lambda$CDM cosmology (orange) and one for the EDE model with $f_{\rm EDE}=0.15$ (blue). In both cases, the posterior constraints are consistent with their respective ground truth latent parameters, thus demonstrating that our pipeline returns unbiased and accurate constraints in latent space. Since the two mock data differ only by the fraction of EDE ($f_{\rm EDE}=0$ in one case and $f_{\rm EDE}=0.15$ in the other), this validation test also shows which latents carry information about this cosmological parameters. We find that nearly all latents carry information about EDE, except for latent 4, 6, 7, 8. This means that the latter affects many (not just one) independent degrees of freedom in the CMB temperature power spectrum. In Sec.~\ref{sec:interpret} we will show that those latents which appear insensitive to the $f_{\rm EDE}$ parameter are largely subdominant in the overall information compared to the other latents, and are therefore responsible for only minimal changes in the CMB spectrum.

To validate our pipeline even further, we run two additional tests considering two mock datasets with `ground truth' 8D latent values which are at the edge of the range covered by the test set cosmologies in latent space. This allows us to test the robustness of the VAE in returning unbiased constraints even when the true spectrum is an unlikely case amongst our test cosmologies. Even in such extreme cases, we find that the latent parameter constraints are unbiased and accurate with respect to the ground truth values, yielding posteriors similar to the case shown in Fig.~\ref{fig:mocktest}, which we do not show for brevity.

\subsubsection{Latent parameter constraints from \rm{Planck} \it{data}}
\label{sec:planck_constraints}
Next, we run our analysis on real data: we compare the $D_{\ell}^{\rm TT}$ theoretical predictions, generated by the VAE decoder from sampled points in latent space, and the \Planck{} data vector for $D_{\ell}^{\rm TT}$. Our analysis in this work will be entirely in latent space; however, we also tested training a neural network to map latent variables to cosmological parameters and reconstruct cosmological constraints, obtaining results in $\sim\,2\sigma$ agreement with direct inference on the cosmological parameters. This transformation confirmed our results from the latent interpretation through mutual information and latent traversals, therefore we decided to omit it for brevity; we further discuss it in Sec.~\ref{sec:conclusions}.

Fig.~\ref{fig:lcdmrefmodel_planck} shows the 1D and 2D marginalized posterior probability distributions for the latent parameters $\bm{z}$ given the \Planck{} data. We present the two VAE cases: one trained on $\Lambda$CDM cosmologies with a 5-dimensional latent space, and one trained on EDE cosmologies with an 8-dimensional latent space. The posterior distributions of the 5D $\Lambda$CDM latent parameters are shown in the left panel of Fig.~\ref{fig:lcdmrefmodel_planck} in orange, and the 8D EDE ones in the right panel in blue. The widths of the contours are mainly driven by the covariance matrix used in the likelihood. Note that the contours widths generally cover down to 2\% of the entire latent parameter space covered by the test set cosmologies, indicating that the latent parameters are very tightly constrained by the data; we show the extent of the posterior constraints compared to the latent space range covered by the entire test set in Appendix~\ref{app:latentpriors}.

We compare the latent posterior constraints against several theoretical expectations. By `theoretical expectation', we mean the range of latent values corresponding to a \textit{single} cosmology i.e.\ a single set of cosmological parameter values. The encoder always returns a distribution in latent space, so even for a single set of cosmological parameters one obtains a distribution in latent space (and hence the 2D contours in Fig.~\ref{fig:lcdmrefmodel_planck}). In the right and left panels of Fig.~\ref{fig:lcdmrefmodel_planck}, we show in green the range of latent values corresponding to a single cosmology (the `theoretical expectations'), that is a $\Lambda$CDM cosmology with cosmological parameters set by the \Planck{} best-fit values. To obtain the green latent distribution, we take the best-fit $\Lambda$CDM cosmological parameters from Ref.~\cite{Planck:2018vyg}, use \texttt{CLASS} to generate $D_{\ell}^{\rm TT}$ and use the trained VAE encoder to map $D_{\ell}^{\rm TT}$ to its encoded $L$-dimensional Gaussian distribution. The comparison shows that our latent posterior constraints are consistent with the latent values corresponding to the best-fit $\Lambda$CDM cosmology from \Planck{} obtained from a traditional Bayesian approach. This is the case for both the $\Lambda$CDM latents and the EDE latents.

In the right panel, we additionally compare our constraints to another theoretical expectation value; again, this is given by the latent values corresponding to a single cosmology. This time, we consider the latent values corresponding to the best-fit EDE cosmology under \texttt{Plik\_lite} TT data (including a \Planck-informed prior on $\tau = 0.0506 \pm 0.0086$ \cite{Planck:2018vyg}). These are shown as pink contours in the right panel of Fig.~\ref{fig:lcdmrefmodel_planck}.
We obtain the best-fit EDE cosmology by running a global minimization with the simulated-annealing minimizer \texttt{pinc}~\cite{Herold:2024enb}, yielding the best-fit values $f_\mathrm{EDE}=0.06$, $\log z_\mathrm{c} = 3.4$, $\theta_\mathrm{i} = 2.4$. We find that the latent posterior constraints are consistent with both the best-fit $\Lambda$CDM cosmology reported by Ref.~\cite{Planck:2018vyg} and the best-fit EDE cosmology under \texttt{Plik\_lite}. Our results are therefore consistent with previous constraints obtained with traditional parameter inference techniques using similar CMB data.\footnote{Our constraints are based on the \texttt{Plik\_lite} TT likelihood, while previous constraints were based on the full \texttt{Plik} likelihood (combined with other data). We used the \texttt{Plik\_lite} TT likelihood because it allows for $D_\ell^\mathrm{TT}$ as input rather than requiring cosmological and nuisance parameters. We verified that \texttt{Plik\_lite} TT gives comparable (albeit slightly looser constraints) on EDE than \texttt{Plik} TT using \texttt{MontePython}~\cite{Audren:2012wb, Brinckmann:2018cvx}.}

Although the two cosmologies (denoted as `theoretical expectations') are mildly separated in latent space, the CMB temperature power spectrum alone is unable to differentiate between those two models, yielding constraints that are consistent with both cosmologies. This is not surprising since EDE was constructed in such a way as to preserve the fit to CMB data while allowing for higher values of $H_0$. In order to probe the ability of EDE to resolve the Hubble tension in latent space, an inclusion of direct measurements of $H_0$ (e.g.\ \cite{Breuval:2024lsv, Murakami:2023xuy, Freedman:2024eph, TDCOSMO:2023hni, Vogl:2024bum}) into the training process is necessary, which is left to future work.

\begin{figure}
\centering
\includegraphics[trim={0.2cm 0.4cm 0.2cm 0.2cm},clip, width=\linewidth]{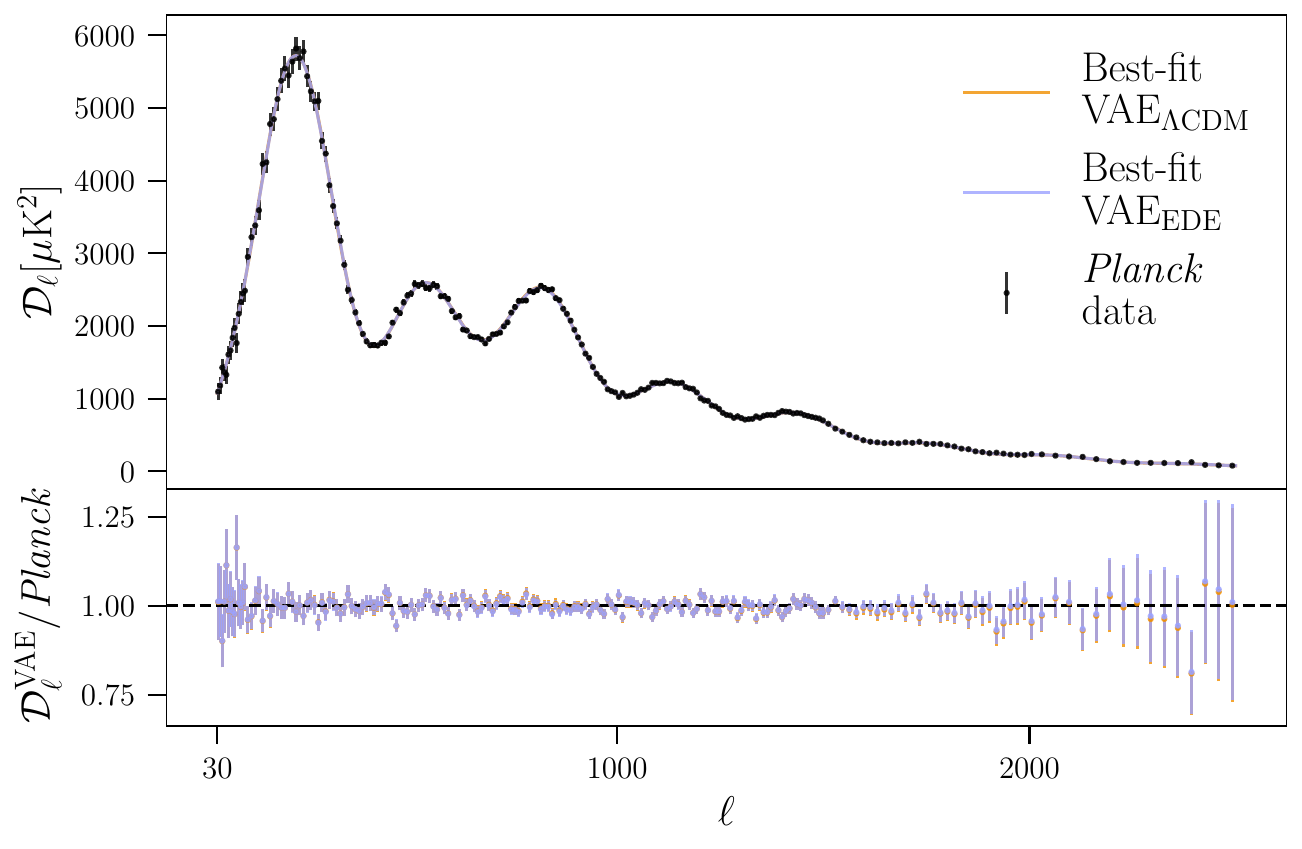}
\caption{Comparison of the measured \Planck{} TT spectrum (black points), the reconstructed \Planck{} best-fit under the $\Lambda$CDM VAE (orange), and the reconstructed \Planck{} best-fit under the EDE VAE (blue), with the ratio between the reconstructed and measured spectra in the bottom panel.}
\label{fig:planck_compare}
\end{figure}

Finally, we reconstruct the CMB temperature power spectrum from the best-fit point in parameter space. We show the best-fit reconstructed spectra from the VAE$_{\Lambda \rm{CDM}}$ and the VAE$_{\rm{EDE}}$ models, compared to the \Planck{} data, in Fig.~\ref{fig:planck_compare}. The VAE models are able to reconstruct the CMB power spectrum at great accuracy throughout the entire $\ell$-range, further validating our approach.

\section{Cosmological information in latent space}\label{sec:interpret}
In this section, we interpret the latent space in terms of the cosmological information in the CMB temperature power spectrum. 
To gain some intuition on the information encoded in the latents, we perform a qualitative analysis where we vary each latent systematically and observe the induced changes in the CMB power spectrum: this is known as a \textit{latent traversal} analysis. We then perform a quantitative analysis by measuring the mutual information (MI) between the latent parameters and the cosmological parameters. We start with introducing the mathematical background of MI and then move on to interpreting the $\Lambda$CDM and EDE latents respectively.

\subsection{Mutual information}
\label{sec:MI_calculation}
MI is a measure of the amount of information shared between two variables $x$ and $y$, given by:
\begin{equation}\label{eq:MI}
\operatorname{MI}\left(x,y\right)=\iint  p(x, y) \ln \left[\frac{p(x, y)}{p(x)\, p(y)}\right] \mathrm{d} x \,\mathrm{d} y \,,
\end{equation}
where $p(x)$, $p(y)$ and $p(x,y)$ are the marginal and joint distributions of $x$ and $y$, respectively.
MI is zero if and only if two variables are statistically independent; we refer the reader to Ref.~\cite{Vergara15} for a complete review. 

We calculate MI using the \mbox{GMM-MI} package \citep{Piras2023},\footnote{\url{https://github.com/dpiras/GMM-MI}} which fits a Gaussian mixture model to the joint distribution of $x$ and $y$ samples to provide a robust estimate of MI along with its associated uncertainty via bootstrapping.
Previous work has already demonstrated the utility of MI in the physical interpretation of latent spaces in the context of predicting the properties of final cosmic structures such as (sub)halo density profiles \citep{Lucie-Smith2022, Lucie-Smith2023, Lucie-Smith:2024xsx} and the halo mass function \citep{Guo_2024}. 

As we show in Sec.~\ref{sec:interpretLCDMlatents} and Sec.~\ref{sec:latentinterpretation}, we use MI to quantify the shared information between each latent parameter and each cosmological parameter; samples from their joint distribution are given by the set of cosmological parameters and latents corresponding to the same spectra. We also use MI to assess the disentanglement of the latent variables in tuning $\beta$ for each VAE: we find that the maximum value of MI between pairs of latents is $\mathcal{O}(10^{-2})$ nat, significantly smaller than the MI between latents and cosmological parameters, thus confirming the disentanglement.

\begin{figure*}
    \centering
    \includegraphics[trim={0.cm 0.5cm 0.35cm 0.2cm},clip,width=0.92\linewidth]{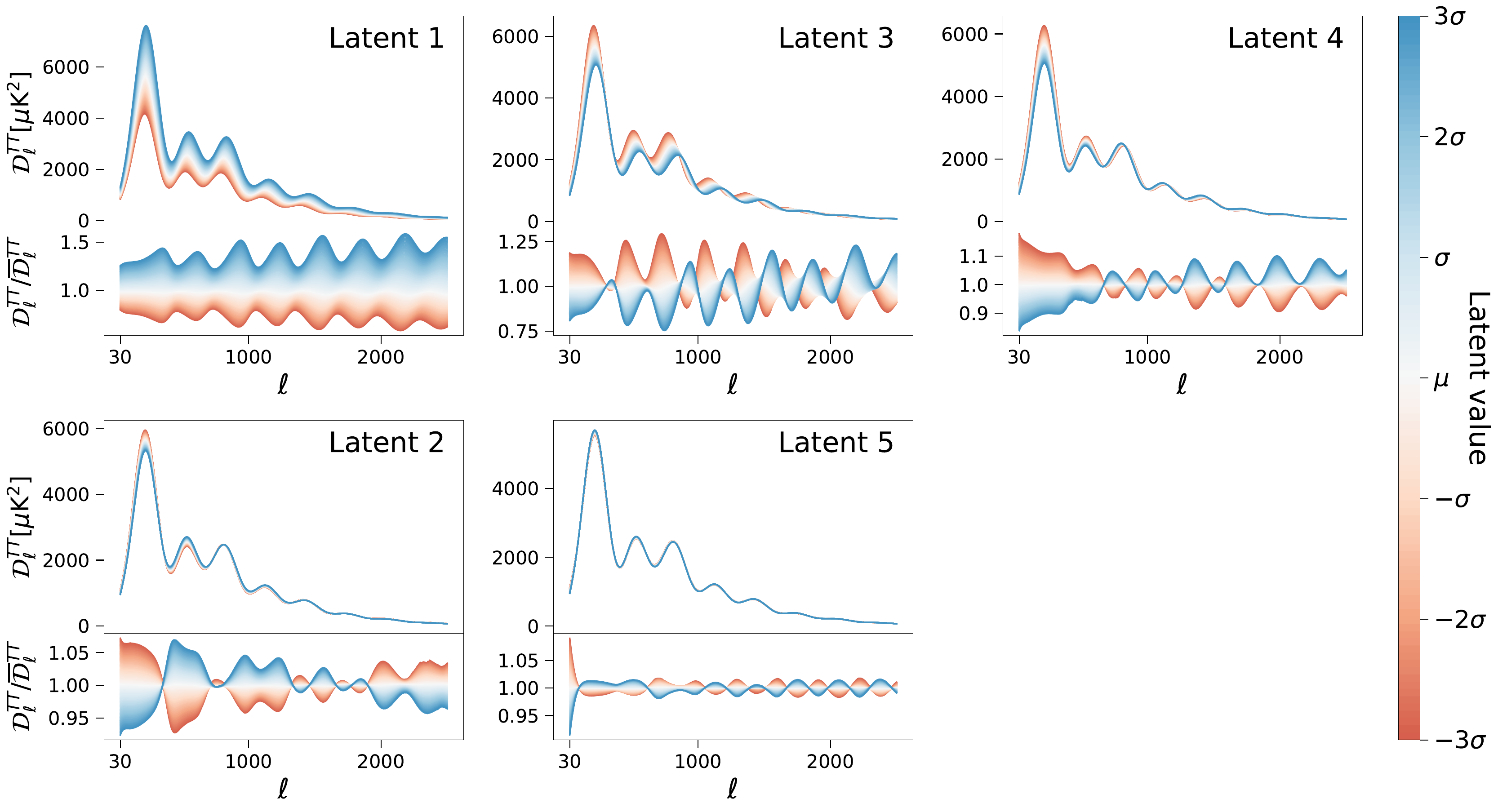}
    \vspace{-0.4cm}
    \caption{Variations in the reconstructed power spectrum when varying one latent of the VAE$_{\Lambda\mathrm{CDM}}$ model systematically, while fixing all others to their mean value across the test set. Each latent is varied within the range [$\mu -3\sigma$, $\mu +3\sigma$], where $\mu$ and $\sigma$ are the mean and standard deviation of the latent across the test set. The bottom panels show the relative change with respect to the mean reconstructed power spectrum, $\overline{\mathcal{D}}_{\ell}^{\rm{TT}}$. The panels are ordered from the most (top left) to the least (bottom right) informative latent.
    }
    \label{fig:lcdm_traversals}
\end{figure*}
\begin{figure}
\centering
\includegraphics[trim={0.cm 0.3cm 0.25cm 0.3cm},clip,width=0.755\columnwidth]{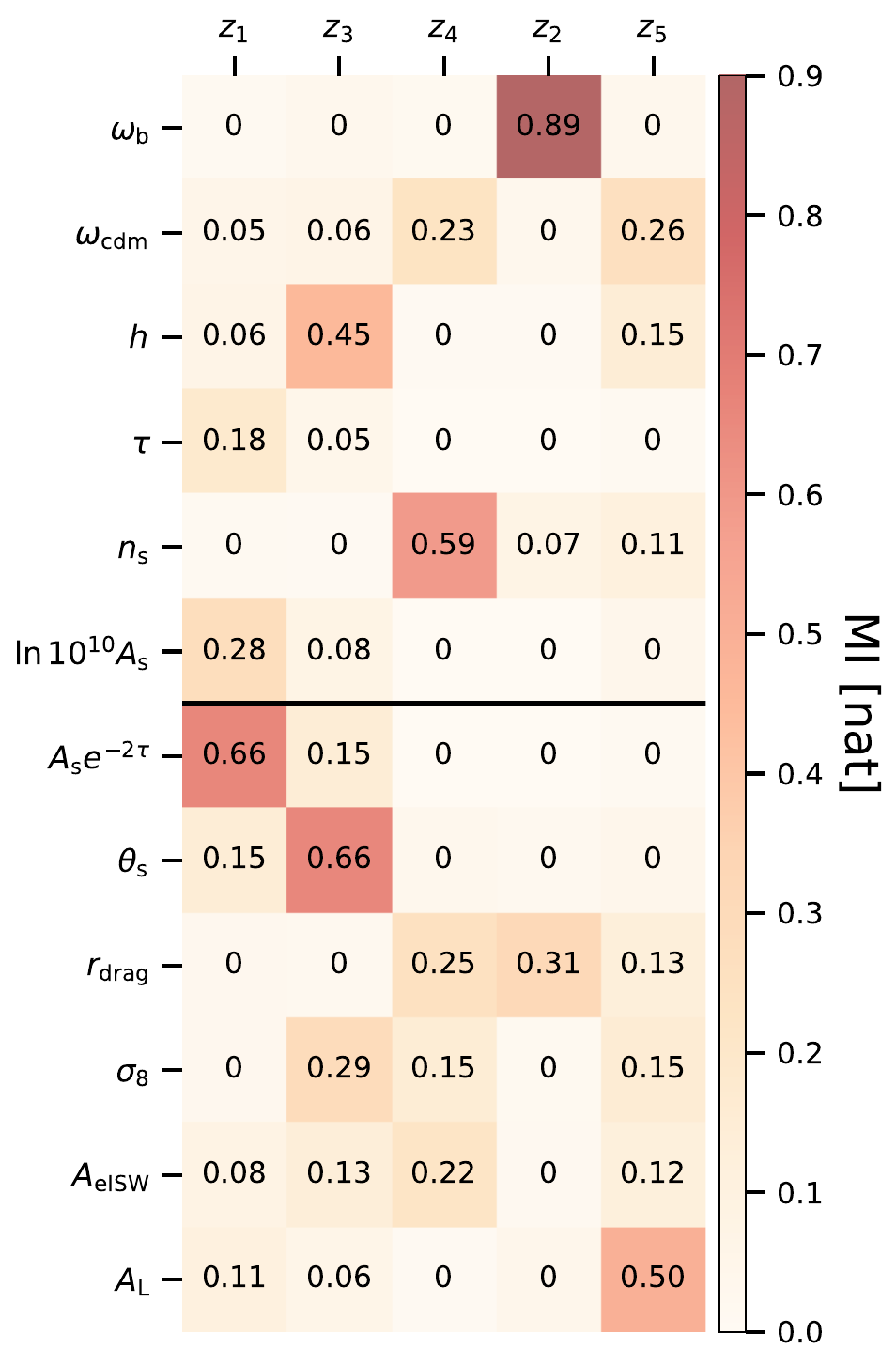}
\vspace{-0.4cm}
\caption{Mutual information (MI) values between the VAE$_{\Lambda \rm CDM}$ latents, $z_1$ to $z_5$, ordered as in Fig.~\ref{fig:lcdm_traversals}, and cosmological parameters (both fundamental and derived, separated by a solid black line). All values below 0.05 nat are indicated as zeros, while MI uncertainties are not reported as they are all small, of order $\mathcal{O}(10^{-3})$ nat.}
\label{fig:MI_values_lcdm}
\end{figure}

\subsection{Interpretation of the VAE$_{\Lambda \mathrm{CDM}}$ latents}
\label{sec:interpretLCDMlatents}
Here, we interpret the latent parameters discovered by the VAE$_{\Lambda\rm{CDM}}$ model which we found to be necessary and sufficient to reconstruct the CMB TT power spectrum in Sec.~\ref{sec:results}.
The latent traversal plots for each of these latents are shown in Fig.~\ref{fig:lcdm_traversals}. In each panel, we show the predicted spectra as we systematically vary the value of one latent, while keeping the others fixed to their mean value. The panels in the top row are ordered from the most (top-left) to the least (bottom-right) informative latent. The latents yield non-trivial modifications to the CMB spectra, including changes to the amplitude, tilt, height and position of the peaks, and more. The induced changes can be compared to well-known physical effects such as the early integrated Sachs-Wolfe (ISW) effect, which is boosted in the context of the EDE model~\cite{Vagnozzi:2021gjh}, and the phenomenological parametrization of the CMB presented in Ref.~\cite{Hu_2001}, as well as to the response of the CMB TT power spectrum to individual cosmological parameters \cite{Komatsu2020}. We show the latter in Appendix~\ref{app:Dl_changes_cosmo_params}, which will be helpful when drawing similarities between the response of the CMB to a cosmological or latent parameter.

Fig.~\ref{fig:MI_values_lcdm} quantifies the shared information between each latent and the fundamental cosmological parameters (top six rows), as well as derived parameters which are more closely related to physical features of the CMB (bottom five rows). The latter include the parameter combination $A_{\rm{s}} \exp{(-2\tau)}$, which determines the amplitude of the CMB TT power spectrum, the angular size of the sound horizon at the time of recombination $\theta_\mathrm{s}$, the sound horizon scale at the baryon drag epoch $r_{\rm{drag}}$, the mass variance of density fluctuation on 8 $\textrm{Mpc}\, h^{-1}$ scales $\sigma_8$, a proxy for the amplitude of the early ISW effect $A_\mathrm{eISW}$, and a proxy for the lensing amplitude $A_{\mathrm{L}}=\max_{\ell} \ell^2 (\ell+1)^2 C_{\ell}^{\varphi\varphi}$, where $C_{\ell}^{\varphi\varphi}$ is the power spectrum of the lensing potential. All these derived parameters are computed with \texttt{CLASS}. To calculate $A_\mathrm{eISW}$, we first compute the contribution of the early ISW effect, which mainly affects the first acoustic peak; $A_\mathrm{eISW}$ is then defined as the maximum early ISW amplitude, namely, $\max_{\ell} C_{\ell}^\mathrm{eISW}$. 

The combination of latent traversals and MI provide us a complementary and thorough understanding of the information content in the latent space. We interpret each of the five latents as follows.

\begin{itemize}
    \item The most informative latent (latent 1) controls the amplitude of the power spectrum: this is parametrized by the combination $A_\mathrm{s}\, \exp({-2\tau})$. This interpretation is confirmed by the high MI between the latter parameter combination and $z_1$. The latent carries lower amounts of information about the individual parameters $\tau$ and $A_{\rm{s}}$, as breaking their degeneracy would require additional polarization power spectra or low-$\ell$ data \cite{Planck:2019nip}.
    Although one might expect a correlation of this amplitude-sensitive latent with $\sigma_{8}$, such correlation is washed out by the dependence of $\sigma_{8}$ on other cosmological parameters, which have little influence on this latent.
    We further note small shifts of the acoustic peaks related to $\theta_\mathrm{s}$ due to this latent.
    
    \item The next latent (latent 3) controls the horizontal position of the acoustic peaks, thus yielding high MI with the angular scale of the sound horizon $\theta_\mathrm{s}$ and the Hubble parameter $h$. This latent is also the one with most MI about $\sigma_8$: since $\sigma_8$ is defined as density fluctuations at a radius of 8 $\textrm{Mpc}\, h^{-1}$, it is correlated with $h$ and thus the MI with the $h$-sensitive latent is not surprising (e.g.\ \cite{Sanchez:2020vvb, Forconi:2025cwp}).
    \item Latent 4 determines the tilt of the power spectrum, parametrized by $n_\mathrm{s}$, mixed with changes in the acoustic peak heights as induced by the amount of cold dark matter, $\omega_\mathrm{cdm}$. Changes in the height of the first few peaks are due to the decay of the potential during the radiation era. This latent is also correlated with the amplitude of the early ISW effect, $A_\mathrm{eISW}$, which additionally contributes to a boost in the height of the first peak and is closely related to $\omega_\mathrm{cdm}$.
    \item Latent 2 has a very clean interpretation as it resembles the response of the CMB power spectrum to $\omega_\mathrm{b}$ alone: we can clearly recognize the distinct even-odd modulation of the acoustic peaks in the latent traversals. This is reflected in the high MI between the parameter and $\omega_\mathrm{b}$. 
    \item Finally, the most subdominant latent (latent 5) mainly captures the smearing effect of the acoustic peaks due to gravitational lensing; this is confirmed by the high MI between the latent and the lensing amplitude (A$_{\rm L}$). CMB lensing is known to constrain parameters such as $\omega_{\rm cdm}$ and $\sigma_{8}$, further explaining a non-negligible MI between the latent and these parameters.
\end{itemize}
Additionally, the sound horizon at the drag epoch, $r_\mathrm{drag}$ shows significant MI information with those latents which are correlated with $\omega_\mathrm{cdm}$ and $\omega_\mathrm{b}$. This is expected since the epoch at which baryons and photons decouple is closely related to the matter content in the universe.

In summary, we find that the VAE$_{\Lambda \rm CDM}$ disentangles the information in the CMB temperature power spectrum into the expected number of degrees of freedom: the overall amplitude ($A_\mathrm{s}\, \exp({-2\tau}$)), the shift in the sound horizon angular scale ($h$), a boost in the height of the acoustic peaks ($\omega_{\rm cdm}$) combined with changes in the power spectrum tilt ($n_\mathrm{s}$), the even-odd modulation of the peaks ($\omega_\mathrm{b}$), and finally changes to the height of the acoustic peaks ($\omega_{\rm cdm}$) to break the degeneracy between peak height and tilt present in the third latent. 
The fact that there are five degrees of freedom out of six cosmological parameters is expected due to the degeneracy between $\tau$ and $A_{\rm s}$.

\begin{figure*}
    \centering
    \includegraphics[trim={0.cm 0.5cm 0.38cm 0.2cm},clip,width=\linewidth]{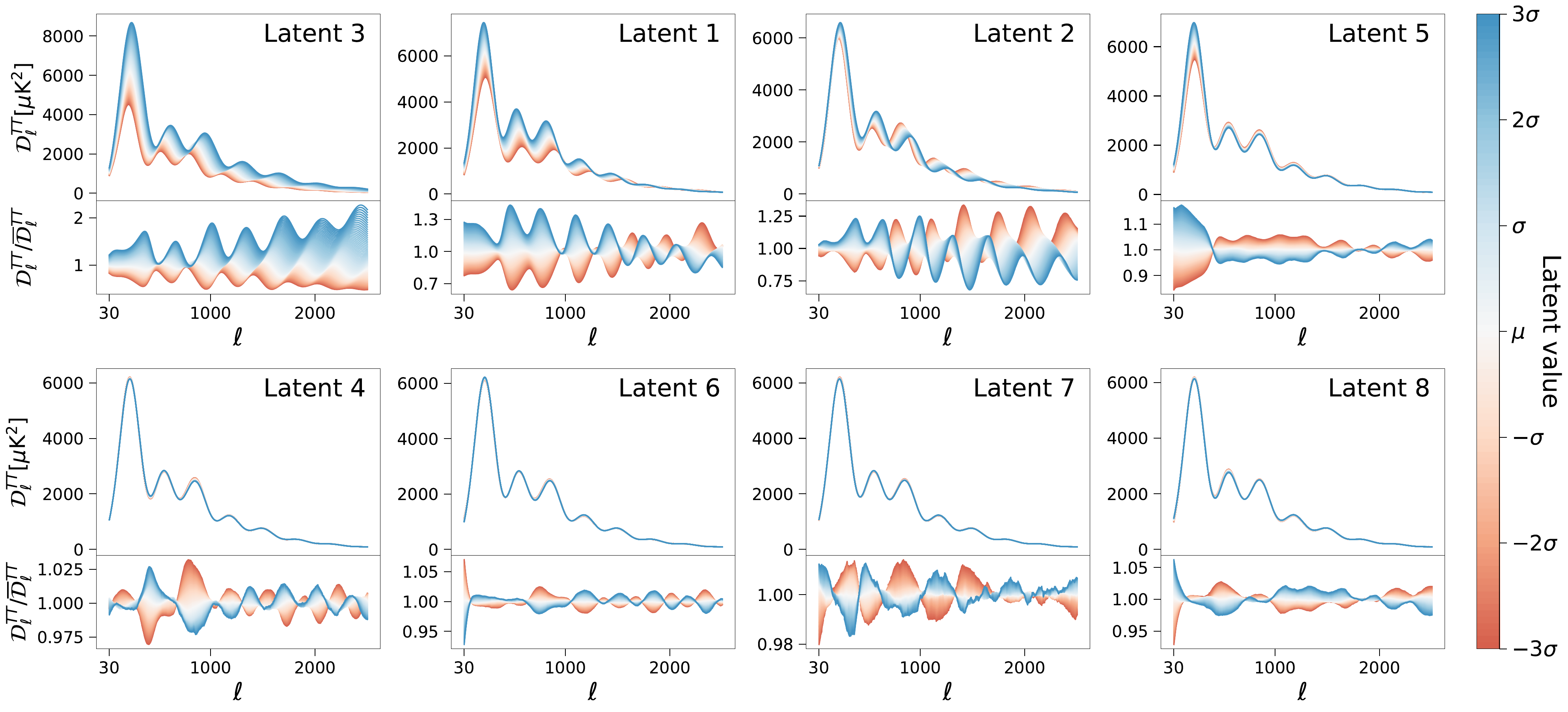}
    \vspace{-0.4cm}
    \caption{Latent traversals for the VAE$_{\rm {EDE}}$ latents, similar to Fig.~\ref{fig:lcdm_traversals}. The panels in the top row are for the dominant latents and are ordered from the most (left) to the least (right) informative one; the latents in the bottom row are the subdominant ones in no particular order.}
    \label{fig:ede_traversals}
\end{figure*}

\begin{figure}
\centering
\includegraphics[trim={0.cm 0.3cm 0.25cm 0.3cm},clip,width=0.96\linewidth]{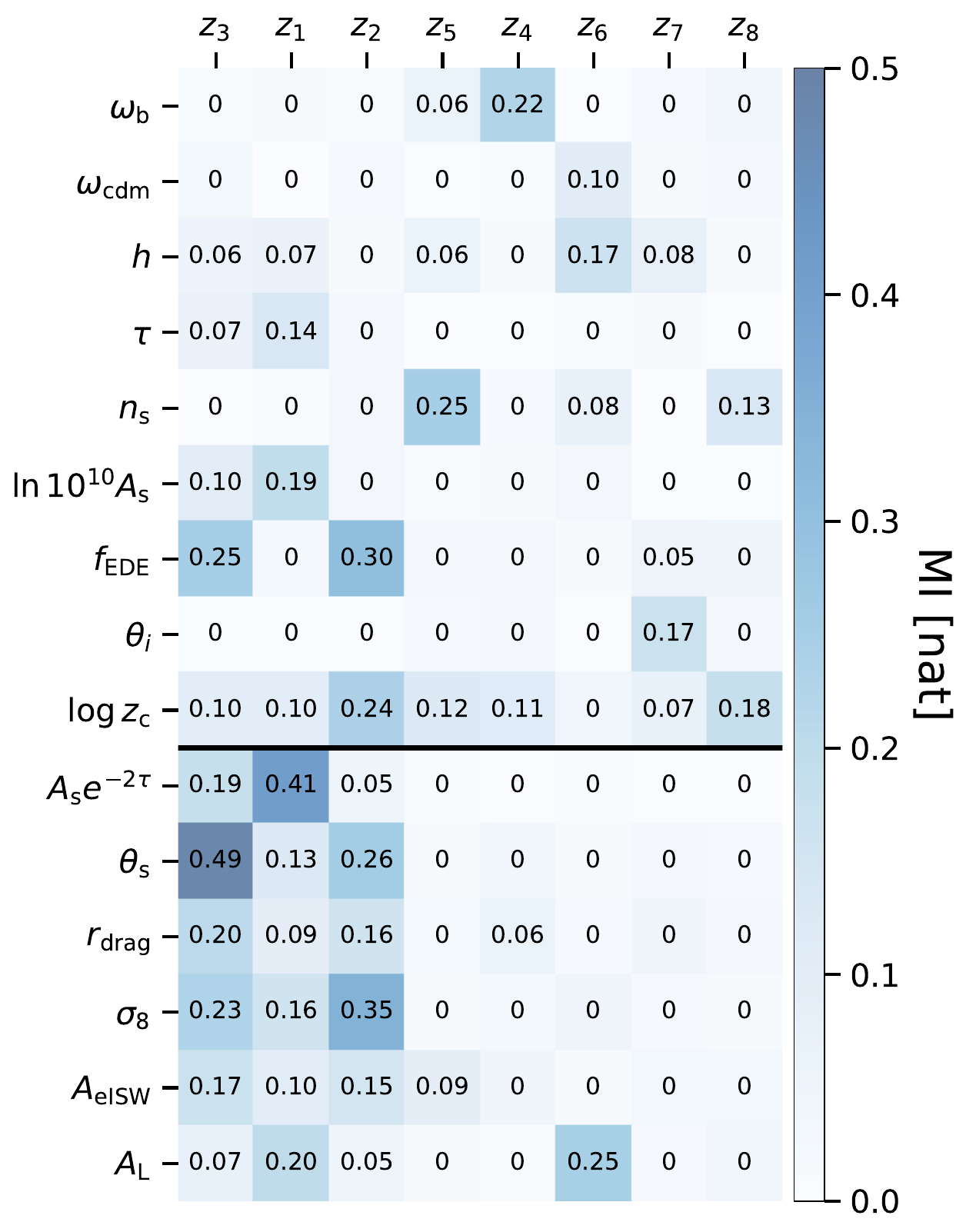}
\vspace{-0.4cm}
\caption{MI between latents of the VAE$_{\rm {EDE}}$ and cosmological parameters, similar to Fig.~\ref{fig:MI_values_lcdm}.}
\label{fig:MI_values_ede}
\end{figure}

\subsection{Interpretation of the VAE$_\mathrm{EDE}$ latents}
\label{sec:latentinterpretation}
We now move on to the less straightforward interpretation of the latents of the VAE$_\mathrm{EDE}$ model, which encode the non-trivial dependency of the CMB temperature power spectrum on the EDE parameters. Fig.~\ref{fig:ede_traversals} shows the latent traversals, similar to the case of $\Lambda$CDM. In this case, the panels in the top row are ordered from the most (left) to the least (right) informative latent; the latents in the bottom row are the subdominant ones in no particular order. The first thing we observe is that there is a hierarchy amongst the latents: latent 3, 1, 2 and 5 induce significant changes (>10\%) in the CMB when varied, meaning that they carry dominant information. Latents 4, 6, 7 and 8 instead induce minor changes that are typically < 5\%; this means that their contribution to the CMB temperature power spectrum is largely subdominant compared to that of the others.

Fig.~\ref{fig:MI_values_ede} quantifies the shared information between each latent and the fundamental cosmological parameters or derived ones. The derived parameters are the same as those used in Fig.~\ref{fig:MI_values_lcdm}, but this time computed for the EDE cosmologies.
We start the interpretation with the dominant latents -- top row of Fig.~\ref{fig:ede_traversals} and first four columns in Fig.~\ref{fig:MI_values_ede}. For comparison, the response of the CMB spectrum to the three EDE parameters can be seen in the bottom row of Fig.~\ref{fig:Dl_cosmo_params} in Appendix~\ref{app:Dl_changes_cosmo_params}.
\begin{itemize}
    \item The most dominant latent is latent 3. It has a combined effect of shifting the sound horizon (primary) and the amplitude of the \Dl{} (secondary). 
    We find a high MI between the latent and the sound horizon $\theta_\mathrm{s}$, the amplitude-related parameters i.e.\ $A_{\rm{s}} \exp{(-2\tau)}$, $\sigma_8$, and the EDE fraction, $f_{\rm{EDE}}$. This latent is the one most sensitive to $A_\mathrm{eISW}$: this is in line with the well-known impact of EDE, which boosts the early ISW effect \cite{Vagnozzi:2021gjh}.
    
    \item The second most dominant latent is latent 1, which carries mostly amplitude information with some small shifts of the acoustic peaks. The changes in amplitude are primarily affected by the well-known combination $A_{\rm{s}}\,\exp{(-2\tau)}$, with hardly any contribution from EDE (in contrast to the previous latent 3). It is interesting that the VAE does not prefer a disentanglement between vertical shift (amplitude) and horizontal shift (sound horizon), but rather disentangles the amplitude effect of $A_{\rm{s}}\,\exp{(-2\tau)}$ present in standard $\Lambda$CDM cosmologies to that of $f_{\rm EDE}$.
    
    \item Latent 2 encodes the unique signature of the impact of EDE on the CMB temperature power spectrum.
    It is in fact primarily correlated to $f_{\rm EDE}$ and the critical redshift $z_{\rm{c}}$, and shares no information with the standard $\Lambda$CDM cosmological parameters. This implies that the VAE was able to isolate the unique effects of EDE which are not correlated with $\Lambda$CDM; these effects include non-trivial changes to the overall amplitude of the power spectrum and the horizon scale. This latent also shows a high MI with $\sigma_8$, confirming the impact that EDE has on $\sigma_8$, which leads to a worsening of the $S_8$ tension \cite{Smith:2019ihp, Hill:2020osr, Ivanov:2020ril, Vagnozzi:2021gjh, Ye:2021nej, Pedreira:2023qqt}.

    \item The next latent in terms of importance is latent 5. We find that this latent captures the effect of a changing slope of the CMB power spectrum as encoded by $n_{\rm{s}}$. The MI between the latent and the physical parameters also confirms that the latent shares a significant amount of information with $n_\mathrm{s}$, and has no information about all parameters.
\end{itemize}

Similar to the $\Lambda$CDM case, four latents contain most of the information in the CMB temperature power spectrum for EDE cosmologies; yet, an additional four second-order latent parameters are required to achieve an accuracy well below the \Planck{} errors.
The subdominant latents induce smaller changes to the CMB, and are thus more difficult to interpret by visual inspection alone. However, the MI gives us a direct measurement of their information content in terms of known parameters, and the comparison with the response of the CMB to cosmological parameters also aids the interpretation. Latent 4 has non-zero MI only with $w_{\rm{b}}$ and $z_{\rm{c}}$: we find that this latent induces an even-odd modulation of the first two peaks and the first trough, in a way that resembles the effect of $w_{\rm{b}}$ at fixed sound horizon. Instead, the high-$\ell$ variations are sensitive to the impact of $z_{\rm{c}}$.
Latent 6 induces small changes in the height and position of the acoustic peaks in a similar fashion to gravitational lensing, which in turn depends also on $\omega_\mathrm{cdm}$ and $h$; this is confirmed by Fig.~\ref{fig:MI_values_ede} which displays in particular a high MI between this latent and the gravitational lensing amplitude $A_{\rm L}$. As opposed to $\Lambda$CDM, there is no strong correlation of the latent controlling $h$ with $\theta_\mathrm{s}$: this might be due to the impact of EDE on the $h$-$\theta_\mathrm{s}$ relation.
Latent 7 and 8 induce shifts of the height and position of the acoustic peaks at the percent level. They show non-zero albeit small MI with some of the EDE-related parameters, as well as $h$ and $n_\mathrm{s}$. Latent 7 is the only latent containing information about the initial value of the EDE scalar field, $\theta_\mathrm{i}$, which has only a small impact on the CMB power spectra. 

In summary, we find that the majority of the information is captured by four latent parameters in both the $\Lambda$CDM and EDE cosmologies. This suggests that, to first order, EDE is largely degenerate with $\Lambda$CDM, with the exception of latent 2. The latter latent serves as a distinctive signature of EDE, influencing the height of the first peaks -- partly due to an enhancement of the eISW effect -- and modifying the tilt of the power spectrum through the $z_{\rm c}$ parameter. On the other hand, $n_\mathrm{s}$ and $\omega_{\rm b}$ are uniquely specified even in the EDE case by two independent latents, while $ \omega_{\rm cdm}$ is traded off for EDE when EDE is introduced. Therefore, an independent determination of $\omega_{\rm cdm}$ is crucial for breaking this degeneracy, as previously pointed out by Refs.~\cite{Poulin:2024ken, Pedrotti:2024kpn}. 

\section{Conclusions}
\label{sec:conclusions}
In this work, we developed a data-driven approach to efficiently compress the CMB temperature power spectra for $\Lambda$CDM and early dark energy (EDE) cosmologies into a minimal set of independent `latent' parameters that capture the information in the underlying data. The latent parameters are automatically identified by a neural network from the data vector itself; they represent the independent degrees of freedom to which the data is sensitive to, and can be interpreted in terms of the physics they capture. Our approach allows us to place constraints on these parameters, in a similar fashion to cosmological parameters, and compare them to the expected latent values of any given cosmology.

We found that the majority of the information in the CMB temperature power spectrum can be encoded in four disentangled latent parameters for both $\Lambda$CDM and EDE cosmologies; however, achieving an accuracy well within observational systematic and statistical uncertainties requires five parameters for $\Lambda$CDM and eight for EDE.
The VAE thus reduces the cosmological parameter space by one parameter in both cases: this is expected since temperature alone can only constrain five out of six $\Lambda$CDM parameters due to the $A_{\rm{s}} - \tau$ degeneracy. The VAE thus recovers the same number of degrees of freedom as in the $\Lambda$CDM parametrization. Our results also imply that the standard EDE parametrization, made of three parameters, cannot be compressed further without compromising the accuracy in the reconstructed spectra.

Utilizing \Planck{} data, we performed Bayesian parameter inference to constrain these physical degrees of freedom. We find that our constraints are in agreement with the expected latent values of a $\Lambda$CDM cosmology and an EDE cosmology with parameters given by Ref.~\cite{Planck:2019nip} and the \texttt{Plik\_lite} best-fit, respectively. This confirms the validity of our approach against previous work in the literature which used the same data and a traditional cosmological parameter inference approach. In particular, we confirm that CMB temperature data alone cannot discriminate between a $\Lambda$CDM cosmology and one with a small amount of early dark energy ($f_{\rm EDE} \approx 0.06$) prior to recombination.

Latent traversals and MI allowed us to physically interpret the latent parametrizations. In the case of the $\Lambda$CDM model, the VAE's five latent parameters have a direct physical interpretation. The two leading latents encode the amplitude and position of the acoustic peaks ($A_\mathrm{s}\exp({-2\tau})$, $\theta_\mathrm{s}$), while a third one the even-odd modulation of the peaks ($\omega_\mathrm{b}$). The fourth latent jointly encode the height of the acoustic peaks and the tilt of the power spectrum ($\omega_\mathrm{cdm}$, $n_\mathrm{s}$), while the last one captures the secondary effect of gravitational lensing.

In the case of EDE, a similar set of latents emerged, although also capturing the influence of EDE in e.g.\ the amplitude or the angular scale of the sound horizon. Most importantly, the VAE discovered a new latent not present in the $\Lambda$CDM case, which entirely isolates EDE effects on the CMB temperature power spectrum from those induced by the $\Lambda$CDM parameters. This latent represents a smoking gun signature of EDE, which cannot be disentangled through a direct inspection of the CMB spectra alone, as the impact of EDE could naively resemble that of $\Lambda$CDM parameters. 
Our method instead achieved one of its original goals of isolating unique physical effects in the data using a data-driven approach.

We focused on performing inference in latent space, rather than cosmological parameter space; however, one could wonder whether there exist advantages in obtaining cosmological parameter constraints from the latent ones. 
When performing such mapping -- from latent to cosmological parameter constraints -- we confirmed that our latent constraints translate into cosmological parameters which agree with standard inference approaches within $\sim\,2\sigma$. Future work will investigate further whether sampling the latents could represent a robust alternative to standard inference methods in cosmological space in the presence of degeneracies and prior volume effects.

Our method is broadly generalizable, enabling us to identify which parameters the data is sensitive to through a data-driven, non-linear approach. Focusing on the \mbox{well-established} cosmological probe of the CMB TT power spectrum allowed us to validate our model and explore its capabilities in a controlled environment. We specifically focused on EDE as it is an example of a phenomenological description of a beyond-standard model of cosmology, which poses challenges related to prior volume effects when performing standard Bayesian analyses. Our methodology also holds promise for compressing other cosmological probes, particularly those related to the late-time Universe, which typically rely on large numbers of correlated parameters. These include, for example, the galaxy power spectrum under the effective field theory of large-scale structure (EFTofLSS, e.g.\ \cite{Baumann:2010tm,Carrasco:2012cv,Senatore:2014via,Senatore:2014eva}), which involves many nuisance parameters that can impact the constraints~\cite{Simon:2022lde, Maus:2023rtr, Holm:2023laa, Donald-McCann:2023kpx}. 
In future work we will investigate whether our method can provide new insights into current cosmological tensions between different datasets by revealing which features of the data are responsible for such discrepancies. This will include the incorporation of additional data vectors, including CMB polarization and late-time probes, to further evaluate the benefits of our approach.

\section*{Author contributions}
\textbf{D.P.}: Methodology; Software; Validation; Formal analysis; Investigation; Visualization; Writing - Original Draft, Review \& Editing.
\textbf{L.H.}: Methodology; Data Curation; Validation; Formal analysis; Investigation; Writing - Original Draft, Review \& Editing.
\textbf{L.L.-S}: Conceptualization; Methodology; Validation \& Interpretation; Supervision; Writing - Original Draft, Review \& Editing.
\textbf{E.K.}: Interpretation; Writing - Review.


\section*{Data availability}
We will make data and materials supporting the results presented in this paper available upon reasonable request. The trained models and an example notebook are available at \href{https://github.com/dpiras/VAExEDE}{https://github.com/dpiras/VAExEDE}.


\section*{Acknowledgments}
LH thanks Graeme Addison and Charles Bennett for helpful discussions. LLS thanks Elisa Ferreira for insightful discussions. DP was supported by the SNF Sinergia grant CRSII5-193826 “AstroSignals: A New Window on the Universe, with the New Generation of Large Radio-Astronomy Facilities”. LH was supported a William H. Miller fellowship.
LLS acknowledges support by the Deutsche Forschungsgemeinschaft (DFG, German Research Foundation) under Germany’s Excellence Strategy – EXC 2121 „Quantum Universe“ – 390833306.
EK was supported in part by the Excellence Cluster ORIGINS which is funded by the Deutsche Forschungsgemeinschaft (DFG, German Research Foundation) under Germany’s Excellence Strategy: Grant No.~EXC-2094 - 390783311. Some computations underlying this work were performed on the Baobab cluster at the University of Geneva, while other parts of this work were performed on the \texttt{freya} cluster maintained by the Max Planck Computing \& Data Facility.

\bibliography{main}


\appendix
\section{Comparison between latent posteriors and priors}
\label{app:latentpriors}

\begin{figure*}
    \centering
    \subfloat[VAE$_{\Lambda\rm{CDM}}$]{%
  \includegraphics[width=\columnwidth]{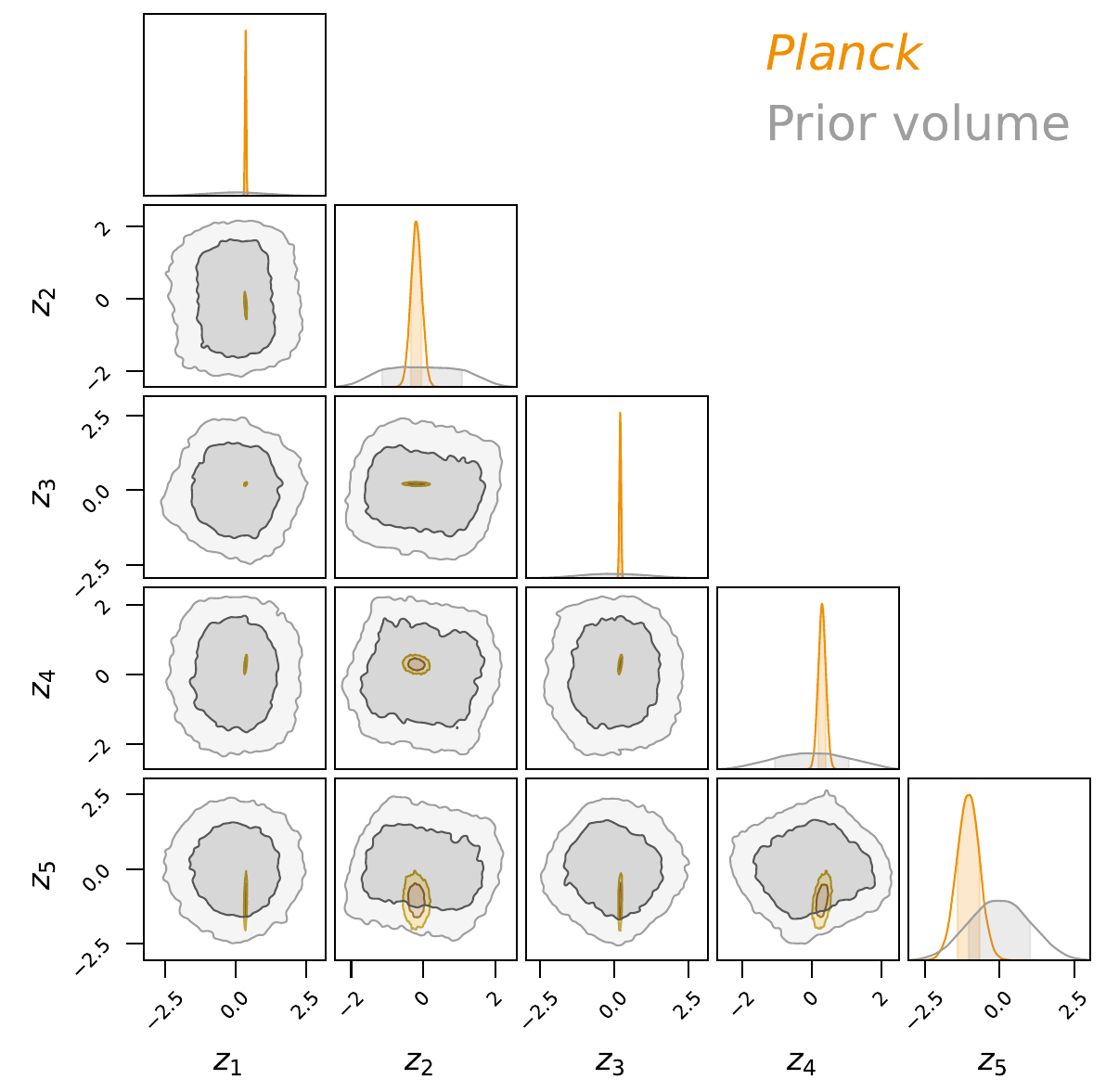}%
}
    \subfloat[VAE$_{\rm{EDE}}$]{%
  \includegraphics[width=\columnwidth]{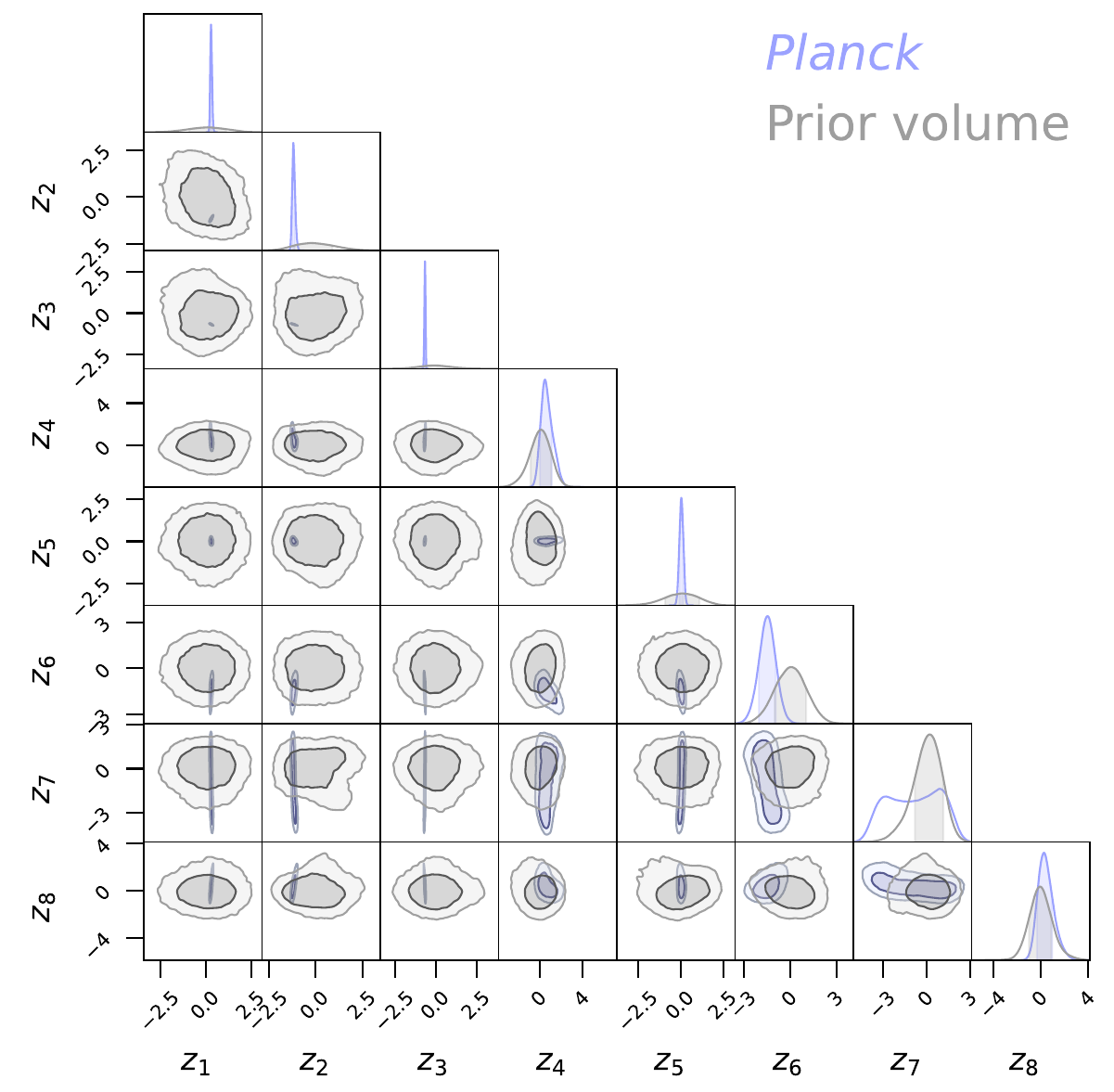}%
}
\caption{Comparison between latent posterior constraints (orange for $\Lambda$CDM on the left, and blue for EDE on the right) and the priors obtained from encoding all test set CMB spectra into the latent space (gray).}
\label{fig:posteriorsvspriors}
\end{figure*}

Fig.~\ref{fig:posteriorsvspriors} compares the latent posterior constraints with the prior volume of the latent space shown as gray contours. The prior contours were generated by encoding all the test set spectra into their respective 5-dimensional (8-dimensional) Gaussian latent distribution predicted by the VAE$_{\Lambda\rm{CDM}}$ (VAE$_{\rm EDE}$) encoder, and sampling from each of those multivariate Gaussians once in order to construct the gray contours. This shows that the latent space is very well constrained compared to the range of possible values of the test set cosmologies. The only exceptions are latent 4, 6, 7, 8 for the EDE case (right panel), which carry very little cosmological information about the spectra (as we demonstrate in Sec.~\ref{sec:interpret}).  As a result, the blue marginalized distributions for those latents are close to their respective prior distributions. In particular, we note that the posterior distribution of latent $z_7$ in the right panel of Fig.~\ref{fig:posteriorsvspriors} even extends beyond the gray distribution. This is because we use broader uniform priors to infer the posterior constraints, and thus there is additional allowed parameter space than what is shown in gray. This behaviour is a result of the poor constraining power of the data for this subdominant latent.

\section{Impact of $\Lambda$CDM and EDE parameters\\on CMB TT spectra}
\label{app:Dl_changes_cosmo_params}
\begin{figure*}
    \centering
    \includegraphics[trim={0.cm 0.5cm 0.35cm 0.2cm}, clip,width=\linewidth]{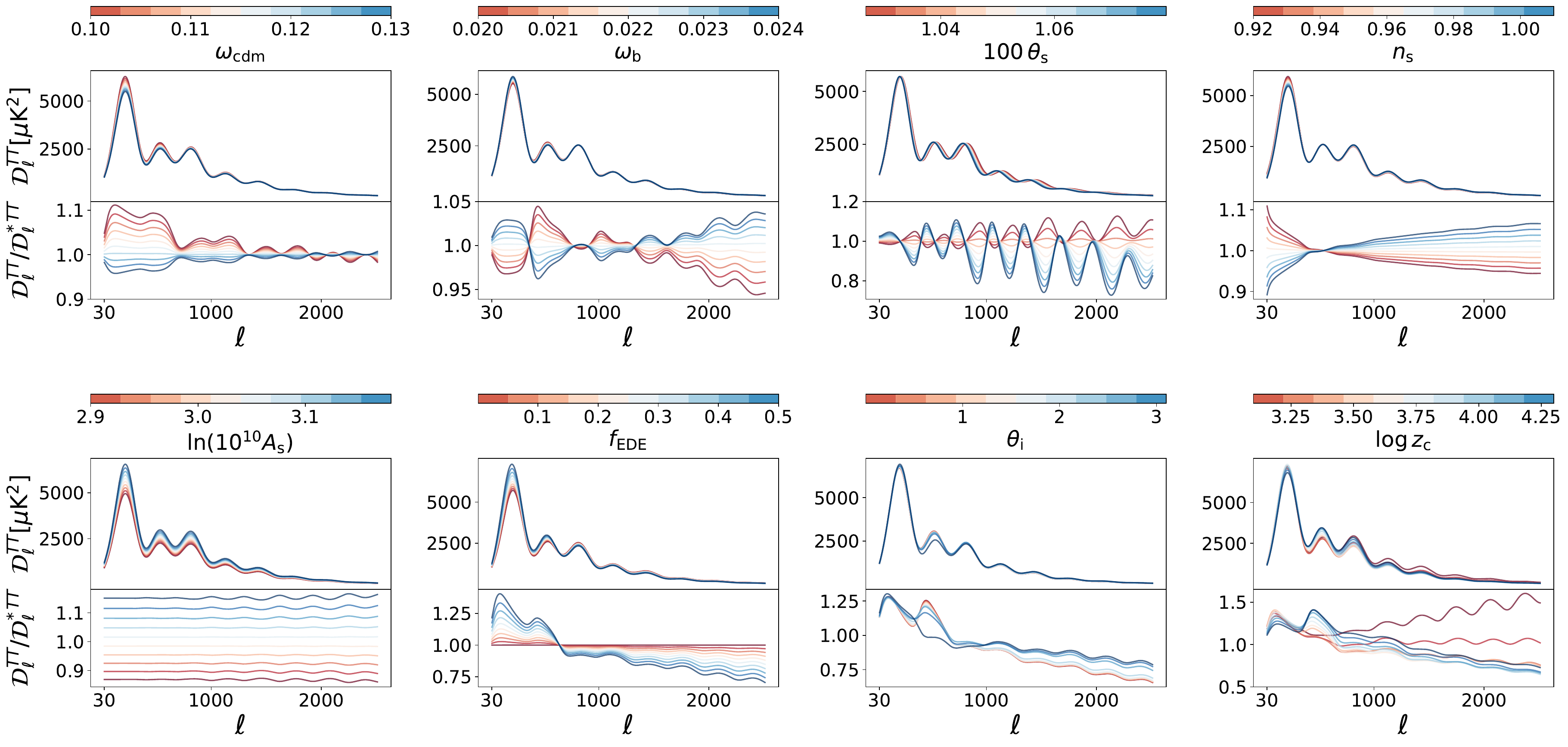}
    \caption{Changes in CMB TT power spectra, $D_\ell^\mathrm{TT}$, induced by varying the conventional $\Lambda$CDM and EDE parameters in the ranges given in Tab.~\ref{tab:prior_lh}. Each parameter is varied one at a time, while keeping all others fixed (including the sound horizon scale). $\mathcal{D}_\ell^\mathrm{*\,TT}$ indicates the spectrum corresponding to the $\Lambda$CDM best-fit values described in Appendix~\ref{app:Dl_changes_cosmo_params}.
    }    \label{fig:Dl_cosmo_params}
\end{figure*}

We show the response of the $D_\ell^\mathrm{TT}$'s to the $\Lambda$CDM and EDE parameters in Fig.~\ref{fig:Dl_cosmo_params} using \texttt{CLASS(\_EDE)} (see also e.g.\ Refs.~\cite{WMAP:2008lyn, Kable:2020hcw}). For each subplot, we fix all other respective cosmological parameters $(\omega_\mathrm{cdm}, \omega_\mathrm{b}, \theta_\mathrm{s}, n_\mathrm{s}, A_\mathrm{s}, f_\mathrm{EDE}, \theta_\mathrm{i}, \log z_\mathrm{c} )$ to the \Planck{} 2018 best-fit values \cite{Planck:2018vyg} for the $\Lambda$CDM parameters and to the best-fit value from Ref.~\cite{Herold:2022iib} (under \Planck{}, BOSS~\cite{BOSS:2016wmc}, and SH0ES~\cite{Riess:2021jrx}, i.e.\ $f_\mathrm{EDE} = 0.13$, $\theta_\mathrm{i} = 2.8$, $\log z_\mathrm{c} = 3.6$) for the EDE parameters. $\mathcal{D}_\ell^\mathrm{*\,TT}$ denotes the spectrum corresponding to the $\Lambda$CDM best-fit values described above. We don't show the impact of varying $\tau$ ($h$) since it is equivalent to the one induced by $A_\mathrm{s}$ ($\theta_\mathrm{s}$). These plots are to be compared to the latent traversals in Sec.~\ref{sec:interpret}.

\end{document}